\documentclass[fleqn,usenatbib]{mnras}

\usepackage{newtxtext,newtxmath}
\usepackage{multicol}

\usepackage[T1]{fontenc}
\usepackage{ae,aecompl}

\usepackage{graphicx}	
\usepackage{amsmath}	

\graphicspath{ {images/} }

\title[PAUS photometric calibration]{The PAU Survey: Photometric Calibration of Narrow Band Images}

\author[Castander et al.]{
F.~J.~Castander$^{1,2}$\thanks{E-mail:fjc@ice.csic.es},
S.~Serrano$^{1,2}$,
M.~Eriksen$^{3}$\thanks{Also at Port d'Informaci\'{o} Cient\'{i}fica (PIC), Campus UAB, C. Albareda s/n, 08193 Bellaterra (Cerdanyola del Vall\`{e}s), Spain},
E.~Gazta\~naga$^{4,1,2}$,
R.~Casas$^{1,2}$,
\newauthor
A.~Alarcon$^{1,2}$,
A.~H.~Bauer$^{1,2}$,
E.~Fern\'andez$^{3}$,
D.~Navarro-Giron\'es$^{1,2}$,
N.~Tonello$^{3}$\footnotemark[2],
\newauthor
L.~Cabayol$^{3}$\footnotemark[2],
J.~Carretero$^{3}$\footnotemark[2],
J.~De Vicente$^{5}$,
J.~Garcia-Bellido$^{6}$,
H.~Hildebrandt$^{7}$,
\newauthor
H.~Hoekstra$^{8}$,
B.~Joachimi$^{9}$,
R.~Miquel$^{3,10}$,
C.~Padilla$^{3}$,
P.~Renard$^{11}$,
E.~Sanchez$^{5}$,
\newauthor
I.~Sevilla-Noarre$^{5}$,
P.~Tallada-Cresp\'{\i}$^{5}$\footnotemark[2]
\newauthor
\\
$^{1}$Institute of Space Sciences (ICE, CSIC), Campus UAB, Carrer de Can Magrans, s/n, 08193 Barcelona, Spain\\
$^{2}$Institut d'Estudis Espacials de Catalunya (IEEC), c/ Esteve Terrades 1, Edifici RDIT, Despatx 212, 08860 Castelldefels, Barcelona, Spain\\
$^{3}$Institut de F\'{i}sica d'Altes Energies (IFAE), Edifici Cn, Campus UAB, 08193, Barcelona, Spain\\
$^{4}$Institute of Cosmology \& Gravitation, University of Portsmouth, Dennis Sciama Building, Burnaby Road, Portsmouth PO1 3FX, UK \\
$^{5}$Centro de Investigaciones Energ\'eticas, Medioambientales y Tecnol\'ogicas (CIEMAT), Avenida Complutense 40, E-28040 Madrid, Spain\\
$^{6}$Instituto de Fisica Teorica UAM/CSIC, Universidad Autonoma de Madrid, Cantoblanco 28049 Madrid, Spain\\
$^{7}$Ruhr University Bochum, Faculty of Physics and Astronomy, Astronomical Institute (AIRUB), German Centre for Cosmological Lensing, \\44780 Bochum, Germany\\
$^{8}$Leiden Observatory, Leiden University, Niels Bohrweg 2,2333 CA, Leiden, the Netherlands \\
$^{9}$Department of Physics and Astronomy, University College London, Gower Street, London WC1E 6BT, UK\\
$^{10}$Instituci\'o Catalana de Recerca i Estudis Avan\c cats (ICREA), 08010 Barcelona, Spain\\
$^{11}$Department of Astronomy, Tsinghua University, Beijing 100084, China
}

\date{Accepted XXX. Received YYY; in original form ZZZ}

\pubyear{2024}

\begin{document}
\label{firstpage}
\pagerange{\pageref{firstpage}--\pageref{lastpage}}
\maketitle

\begin{abstract}
The Physics of the Accelerating Universe (PAU) camera is an optical narrow band and broad band imaging instrument mounted at the prime focus of the William Herschel Telescope. We describe the image calibration procedure of the PAU Survey data. We rely on an external photometric catalogue to calibrate our narrow band data using stars that have been observed by both datasets. We fit stellar templates to the stellar broad band photometry of the Sloan Digital Sky Survey and synthesise narrow band photometry that we compare to the PAUS narrow band data to determine their calibration. Consequently, the PAUS data are in the AB system as inherited from its reference calibrator. We do several tests to check the performance of the calibration. We find it self-consistent when comparing repeated observations of the same objects, with a good overall accuracy to the AB system which we estimate to be at the 2\% precision level and no significant trends as a function of narrow band filter or wavelength. Repeated observations allow us to build a spatial map of the illumination pattern of the system. We also check the wavelength dependence of the calibration comparing to stellar spectra. We find that using only blue stars reduces the effects of variations in the stellar template fitting to broad-band colours, improving the overall precision of the calibration to around 1\% and its wavelength uniformity. The photometric redshift performance obtained with the PAUS data attests to the validity of our calibration to reach the PAUS science goals.

\end{abstract}




\section{Introduction}

The understanding of the Universe requires sampling a large volume to have enough statistical power to reliably measure its structure and evolution. The drive to explore the universe has pushed the development of wide-field cameras to cover large areas of the sky efficiently. Such cameras have allowed the astronomical community to carry out large imaging surveys. In order to efficiently exploit these surveys, they need to be well calibrated. As an example, the determination of the galaxy luminosity function requires a consistent measurement of the flux of the galaxies throughout the volume used to determine it. Moreover, in order to compute the luminosities from the measured fluxes one needs to determine the redshift of the galaxies which, if based on photometric measurements, requires a good calibration of the photometry not to bias the photometric redshift estimates. The same is true for the determination of cosmological observables, like the galaxy two-point correlation function or the weak lensing statistics, where inhomogeneities in the sample selection due to calibration errors can mimic the signal to be measured and produce biased results.

Current wide-area surveys are taken with cameras that have large fields of view. They are equipped with pixelised detectors, normally Charge Coupled Devices (CCDs) in the optical wavelength range. Disregarding the details of the detection process in these devices, the photons received from an object will produce a signal in the detector that will be converted to a digital value (ADU: Analog Digital Unit). The factor to convert from the detected ADUs per unit time (or ${\rm e}^-/s$ if the gain correction is applied) to the flux of the object is known as the zero point (ZP). Normally the ZP is used as an additive term when converting the logarithm of the detected count rate to the object magnitude. The photometric calibration of an astronomical image consists of determining this ZP conversion factor. 

There are several effects that need to be considered when determining the zero point. First, the atmosphere absorbs part of the light in a wavelength-dependent way. The amount of absorption will depend on how long the light travels through the atmosphere. The distance traveled in units of equivalent atmospheres is known as airmass. Then, the light reflects and/or transmits in the telescope and camera optics with a certain efficiency. Later, it goes through a camera filter that has a given wavelength dependent transmission. Finally, the light is collected in the detector with a certain efficiency known as quantum efficiency. Traditionally, the effect of the atmosphere is not included in the zero point, although sometimes it is. As we have remarked in all the previous steps, the ZP is wavelength dependent. It is also position dependent within the focal plane and within the detector area as the optical elements and the detectors are not completely uniform in their transmission and response. 

The calibration of these effects that influence the conversion of photon flux above the atmosphere into observed count rates is difficult to estimate directly and normally is obtained from observations of standard objects whose spectral energy distribution (SED) is known. The knowledge of these SEDs can either come from calibrated spectroscopy or calibrated fluxes at given filters with well defined transmission curves. 

All imaging instruments will have a particular filter transmission system.  Therefore, one needs to apply conversion factors between the filter system of the standards and the observations to calibrate them. These are normally referred to as colour terms. The large field of view instruments will have transmission variations within their large focal planes and thus these colour terms can also be position dependent.

As in any measurement process, the calibration of astronomical images also depends on factors inherent to the measurement process of the objects used to calibrate. Generally, the light from an object is distributed across several pixels due to atmospheric turbulence, the system optics and the intrinsic size and shape of the object compared to the pixel size. The measurement method, either aperture photometry, model fitting or any of its variants, will have an uncertainty due to the Poisson noise, the read-out noise of the detectors and biases from light that may have been neglected in the measurement. In order to minimise uncertainties, stars are generally used as calibration sources as they are point-sources and have normally well understood SEDs, minimising measurement and wavelength-dependent errors.

A night can be considered photometric if the atmospheric extinction correlates with airmass. Traditionally, astronomical images have been calibrated in these photometric nights by taking observations of standard stars~\citep[e.g.,][]{Landolt:92,Landolt:09,Landolt:13,Stetson:00} at different values of airmass to estimate the extinction coefficient and the zero point of the night. With these values one could calibrate the flux of the objects as if they had been observed outside the atmosphere. However, for large multi-band surveys this method is inefficient or impossible to be applied as, for instance, in order to be efficient, observations are taken irrespectively of the photometric conditions of the observing nights. 
Therefore, large surveys need to rely on other methods for their calibration.

The Sloan Digital Sky Survey (SDSS; \citealt{York:00}) represented a major step forward in large area surveys. SDSS built a telescope with an imaging camera~\citep{Gunn:06} and two spectrographs~\citep{Smee:13} to efficiently sample a large fraction of the Northern sky. SDSS designed a new filter system~\citep{Fukugita:96,Doi:10} that has become a reference ever since. The system had no previous calibration and the SDSS devised a new procedure to calibrate their photometry~\citep{Tucker:06}. A network of primary photometric standards was defined with the US Naval Observatory (USNO) 1.0-m telescope.
A purpose-built 0.5 m Photometric Telescope observed the primary standards to define the extinction coefficients and zero points for each observing night and calibrate secondary standards that were then used to calibrate the images taken with the SDSS 2.5-m telescope. This scheme still relied on the transparency of the nights. SDSS chose to take images in photometric nights and spectroscopy when the imaging camera was not taking images. In this way, the nightly calibration could be applied.

In contrast to SDSS, most imaging surveys do not have auxiliary telescopes to calibrate their photometry. Moreover, pure imaging surveys need to be able to observe in non-photometric conditions and still be able to calibrate their observations to be efficient. In order to perform the calibration, the observed objects need to be compared to a reference set that serves as standard. One can use previously calibrated surveys if observations are taken in the same area. The calibration is then set to this reference survey. In this case, one needs to compute the appropriate colour terms between the filter systems. Recently, the {\it Gaia}\footnote{\url{https://sci.esa.int/web/gaia}} space mission has become the main cross-calibration reference system given its all-sky coverage and exquisite photometric accuracy and precision~\citep{Prusti:16,Evans:18,Riello:21}. Another approach is to set up a sparse sample of standards in the same area of the main survey and use it for the calibration, as the Dark Energy Survey\footnote{\url{https://www.darkenergysurvey.org}} (DES, \citealt{DES:05,DES:16}) did for its first year data release~\citep{Abbott:18,Drlica-Wagner:18}.

One can also use specific features of the observed objects that are well located in colour space. Stars occupy a well defined location in colour space~\citep[e.g.,][]{Gaidos:93,Covey:07,Davenport:14}. This stellar locus can be used to improve the photometric calibration once a given normalisation is obtained in one band ~\citep[e.g.,][]{Ivezic:04,deJong:17}.

Another approach was introduced with the analysis of the DES Year 3 data~\citep{Sevilla-Noarbe:21}. They developed the Forward Global Calibration Method \citep[FGCM,][]{Burke:18} that calibrated the entire survey using a forward-modelling approach that incorporates atmospheric and instrumental behaviour, obtaining the best-fit parameters of such a model. FGCM has been incorporated into the Vera C. Rubin Observatory's Legacy Survey of Space and Time (Rubin-LSST; \citealt{Ivezic:19}) and the Hyper Suprime-Cam Subaru Strategic Program Survey (HSC-SSP; \citealt{Aihara:18}) pipelines for use with their datasets.

In surveys where there is some redundancy of the data one can use the different observations of the same objects to uniformly calibrate the survey~\citep[e.g.,][]{Maddox:90,Padmanabhan:08}. This is particularly important for wide area surveys whose observations are taken with many individual exposures through a long time expand.

The wavelength resolution of the calibration of the standards needs to be better than or similar to the resolution of the objects to be calibrated for the calibration process to be effective.   
So far we have considered the calibration of wide area surveys taken in a set of broad band filters (e.g., SDSS; the Kilo Degree Survey\footnote{\url{https://kids.strw.leidenuniv.nl/index.php}}~\citep[KiDS,][]{deJong:13}; DES).
In this case, the standards can also have calibrated photometry in broad-band filters and the calibration process needs to determine the corresponding colour terms. In the case of narrow-band filters, the wavelength resolution of the calibrating standards has to be comparable or better than the wavelength width of the filter transmissions. There is not a standard set of narrow band filters and each instrument has its own filter system. This specificity of the filters forces each instrument to generate its own set of photometric standards, which is a laborious process. Instead, one can use spectrophotometric standards which spectral energy distributions have been calibrated in flux and that have enough spectroscopic wavelength resolution to synthesize the magnitudes of the standards in the desired filter system. Unfortunately, there are very few well-measured spectrophotometric standards. Therefore, narrow band surveys wanting to cover large areas and observe in a variety of photometric conditions need to rely on other objects with well enough known properties as to be able to estimate its narrow band fluxes. 

\hfill

The Physics of the Accelerating Universe  Survey (PAUS) is a project to cover a wide area of the sky with narrow band filters using the PAU Camera \citep[PAUCam,][]{Padilla:2019} to sample the spectral energy distribution of the observed galaxies and obtain their photometric redshift with better accuracy than with broad band imaging. PAUCam is mounted at the prime focus of the 4m-class William Herschel Telescope at the Observatory of ``El Roque de los Muchachos" in the island of La Palma in Spain.  PAUCam is equipped with six broad-band filters ($u$, $g$, $r$, $i$, $z$ and $Y$) with the same design as the DES filters, and 40 narrow-band filters. The narrow band filter system covers the 4500 to 8500 {\AA} wavelength range with 40 filters of 130 {\AA} width separated in wavelength by 100 {\AA} (see \citealt{Padilla:2019} for details). The PAU Survey chose to target the CFHTLS fields~\footnote{\url{https://www.cfht.hawaii.edu/Science/CFHLS/cfhtlsdeepwidefields.html}} where galaxy shape measurements obtained from deep imaging were already available \citep{Heymans:2012,Erben:2013}. The combination of the lensing measurements of the source galaxies with the photometric redshifts of the lensing galaxies can enable a variety of studies \citep[e.g.,][]{Gaztanaga:2012}. The PAUS collaboration has developed a full pipeline to reduce and analyse the images~\citep{Tonello:2019, Serrano:2022} and has demonstrated that accurate photometric redshifts can be obtained with the PAUS narrow band data  (\citealt{Eriksen:2019,Eriksen:2020,Alarcon:2021,Soo:2021,Navarro-Girones:23}). One key step in the pipeline reduction is the photometric calibration of the data that allow us to properly determine the SED of the galaxies and to be able to obtain accurate photometric redshifts and derive the physical properties of the galaxies.

The PAU survey has already published results on data analysis techniques \citep{Cabayol:2019,Cabayol:2021,Cabayol:2023}, narrow band data simulations \citep{Stothert:2018,Tortorelli:2018}, galaxy pairs analysis~\citep{Gonzalez:2023}, Lyman alpha intensity mapping predictions \citep{Renard:2021}, galaxy spectral features and physical parameters \citep{Tortorelli:2021,Renard:2022} and intrinsic alignments \citep{Johnston:2021}. This paper is intended as a reference to support these results explaining the choices adopted to calibrate the narrow band data. In Section~\ref{section: Photometric calibration}, we describe the calibration methodology we have followed to calibrate the PAU survey. In Section~\ref{section: Validation and performance}, we describe the tests carried out to validate the calibration and check its performance. We end up by summarising our findings and presenting our conclusions in Section~\ref{section: Summary and Conclusions}.

\section{Photometric calibration}
\label{section: Photometric calibration}

\subsection{General procedure}
\label{subsec:general_procedure}

In a photometric night, the extinction correlates linearly with airmass. The change in magnitude of an object observed at a given airmass ($m_{\rm obs}$) compared to what would be observed if there was no atmosphere ($m_{\rm o}$) is 
\begin{equation}
\label{eq_airmass}
m_{\rm obs} - m_{\rm o} = k(\lambda)\, X 
\end{equation}
where $k(\lambda)$ is the extinction coefficient that depends on wavelength and $X$ the airmass. The extinction coefficient can be calibrated by taking observations of standard stars of known flux/magnitude at different values of the airmass. Equation~\ref{eq_airmass} should also include a colour term if the filter in which the standard stars are calibrated and the filter to calibrate are different.

This calibration scheme is valid when the night is photometric. However, most current large area imaging surveys need to observe even if conditions are not photometric in order to be efficient.  In this case, one needs to use another calibration procedure.

Observations with CCD detectors measure fluxes in electrons/s (or ADU/s) units. One would like to convert this instrumental flux into a calibrated flux or magnitude. As mentioned in the introduction, this conversion factor is referred to as zero point. Equation~\ref{eq_ZP} shows this relation to compute calibrated magnitudes from the observed flux using the calibration zero point either including the atmospheric extinction term (ZP$_1$) or not (ZP$_2$).   

\begin{equation}
\label{eq_ZP}
\begin{aligned}
m_{\rm obs} & =  -2.5 \, \log_{10} f_{\rm obs} + {\rm ZP}_1  \\ 
        & =  -2.5 \, \log_{10} f_{\rm o} + k(\lambda) \, X + {\rm ZP}_2 
\end{aligned}
\end{equation}
where the fluxes ($f_{\rm obs}$ is the observed flux and $f_{\rm 0}$ is the flux outside the atmosphere) are normally given in units of electrons per second, $e^-/s$.

The PAU Survey data cannot be calibrated in the standard way (e.g., Eq.~\ref{eq_airmass}) because we observe regardless whether the conditions are photometric or not. Moreover, the PAU filters system is unique and there are no photometric standards calibrated in our photometric system. Our procedure then relies on calibrating against objects already calibrated by other surveys in the same fields that have spectral energy distributions that are known or that can be inferred. This strategy is not straightforward as the PAU narrow-band filter system is completely different to any broad band calibration system available. We need to circumvent this problem by modeling the spectral energy distribution of the objects we use as calibrators and applying our knowledge of the PAUCam system response.

Our procedure consists of matching the stars we detect in our observations to stars measured in the Sloan Digital Sky Survey. We prefer to use stars as their photometry is simpler, given that they are point sources and their spectral energy distributions are simpler than those of galaxies or other extragalactic objects. We fit stellar templates to the broad-band SDSS data and use those templates to compute the expected PAU magnitudes using the measured transmission of the PAUCam filter system. Then, We compare the observed magnitudes to the synthetic ones to obtain the zero point calibration of the image. In this way, our zero point includes the extinction coefficient (ZP$_1$ in Eq.~\ref{eq_ZP}) and by definition does not include classical colour terms. Although, our procedure of synthesizing magnitudes in the PAUCam filter system is equivalent to the computation of colour terms. 

Ideally, one would like to use as calibration sources stars with smooth SEDs to avoid the uncertainty of modelling the variations not resolved with the SED sampling provided by the the SDSS broad bands. In that respect, bluer stars are preferred as calibrators compared to redder stars.

We also compute the overall response of the PAUCam filter system using spectrophotometric standards. These stars have been observed in all the PAU filters during twilight throughout our observing campaigns. Unfortunately, this calibration cannot be directly applied to our PAUS data as we have observed in non-photometric conditions and even when observing in photometric conditions, we have not calibrated the extinction coefficient as we did not sample the standards at different values of the airmass due to time limitations. Nevertheless, we can use this overall system calibration to statistically cross-check against the direct calibration of our individual exposures.

Another consideration to take into account when calibrating against observed stars is whether the measured flux of these stars has been corrected or not for atmospheric telluric absorption features~\citep[e.g.,][]{Gullikson:14,Smette:15}. As we shall see below, these absorption features generate uncertainty in the calibration at their wavelengths given their intrinsic variability. 

\subsection{SDSS data}

\begin{figure}
  \centering
  \includegraphics[width=\linewidth]{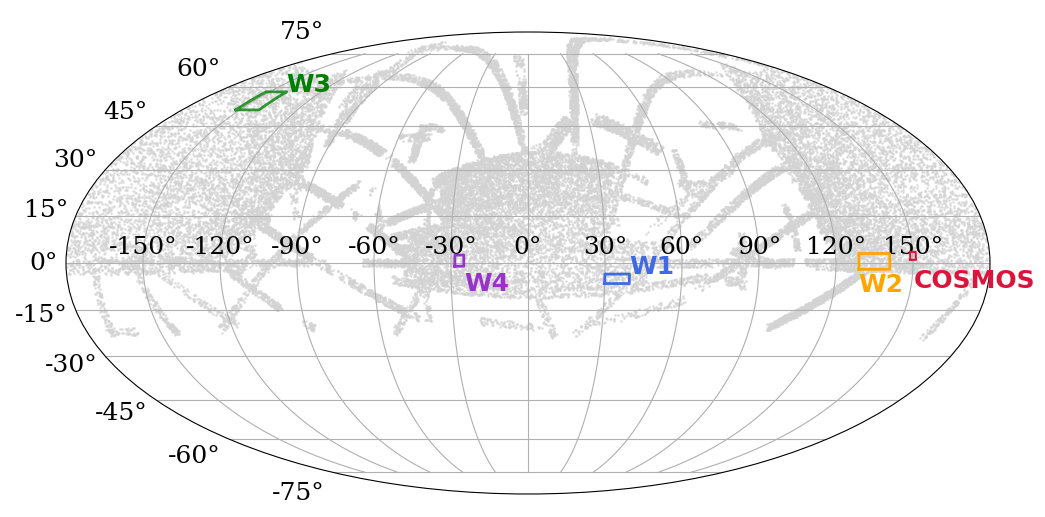}
  \caption{The SDSS DR12 star sample used as reference for the photometric calibration of PAU. This figure only displays around 40\,000 stars brighter than $i_{\rm AB} = 10$, from the total of 260 million stars. The figure also displays the location of the PAUS fields that are within the SDSS footprint. The SDSS stellar density is large enough to have enough stars to perform the PAUS calibration analysis.}
  \label{fig:sdss_stars}
\end{figure}

The PAU Survey has taken data in Northern and Equatorial fields (the W1, W2, W3 and W4 CFHTLS fields and the COSMOS field, see Fig.~\ref{fig:sdss_stars}). At the time the PAU Survey started, the best large area multi-band well-calibrated survey that overlapped with PAUS was the Sloan Digital Sky Survey. SDSS is a large survey observing most of the northern sky with imaging and spectroscopy. 
It has accurate and uniform photometric calibration down to 1\% thanks to their continuous drift-scanning observing strategy and an excellent absolute ubercalibration \citep{sdsscal-Padmanabhan2008}. Therefore, it serves as a good reference for PAUS. 
Ideally we would use the SDSS stellar spectra as spectrophotometric standards to calibrate the PAUS narrow band images. However, there are not enough SDSS stellar spectra to calibrate every single PAUS exposure in all the fields. Therefore we use the stellar photometric sample, composed of $\sim260$ million stars covering 14,500 deg$^2$ of the north sky. The overlap between the PAUS fields and the SDSS DR12 stars is shown in Fig.~\ref{fig:sdss_stars}.

We want to match our point-like source detections to the SDSS star catalogue. We downloaded a copy of the `Star view' table from SDSS which essentially contains the measured photometric parameters for all primary point-like objects, including quasars. From the long list of parameters available, we use only the Point-Spread-Function (PSF) magnitudes (psfMag) in their $ugriz$ broad band set. The PSF measurements provide the highest signal-to-noise measurements of point-like objects such as the stars we want to use as calibration sources. Additionally, there is a \textit{clean} parameter flag that allows us to filter possible photometric problems in the SDSS measurements. We apply a colour cut to the sample to reject a substantial part of the quasar (QSO) population. Based on the colour distributions presented in \citet{Richards:2001}, we apply the following cuts to reject the majority of the quasars and minimise the objects discarded that belong to the main stellar locus:
\begin{align}
\label{eq:colour_cut_qso1}
(g-r) & > 1.25 - 1.12\,(u-g) \\
\label{eq:colour_cut_qso2}
(g-r) & > -0.10 + 0.33\,(u-g)
\end{align}

\begin{figure}
  \centering
  \includegraphics[width=\linewidth]{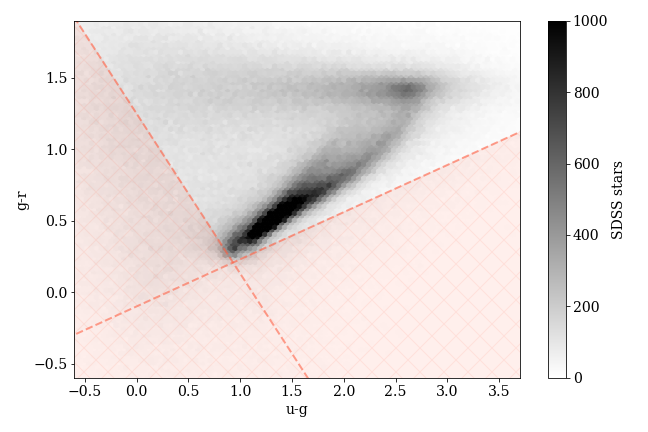}
  \caption{Colour-colour (g-r).vs.(u-g) diagram of the star catalogue in SDSS DR12. This catalogue is made up of point-like objects and therefore QSOs are also included. The proposed cut (Eq.~\ref{eq:colour_cut_qso1} \&~\ref{eq:colour_cut_qso2}, marked as a red shaded region) shows the location populated by QSOs sources that we reject from the sample. This selection eliminates 25\% of the point-like sources.}
  \label{fig:qso_rejection}
\end{figure}

The $ug$-$gr$ colour-colour diagram of the SDSS star sample is shown in Fig.~\ref{fig:qso_rejection}, with the QSO cut we applied. As a result, 25\% of the sample is rejected, leaving a cleaner set of stars for the calibration process.

\subsection{Stellar libraries}

Our method relies on fitting stellar templates to the observed SDSS photometry and using those templates to synthesise the expected PAUS magnitudes. Consequently, we need a comprehensive stellar template sample that properly covers the stellar locus, providing good matches to all the colour combinations observed in the SDSS $ugriz$ set.

\begin{figure}
  \centering
  \includegraphics[width=\linewidth]{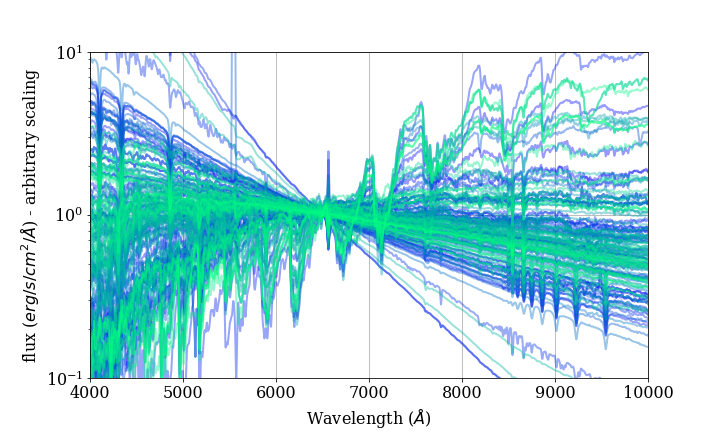}
  \caption{A random subset of 170 stellar templates from the XSL DR2 set~\citep{xsl-gonneau2020} in the wavelength range of interest from 4000 {\AA} to 1 $\mu m$. The y-axis is in units of flux density (erg/s/cm$^2$/{\AA}), normalised at 6500 {\AA}. The colour code is arbitrary.}
  \label{fig:xsl_seds}
\end{figure}

For calibrating the PAUS photometry, we use the X-shooter Spectral Library \citep[XSL,][]{Chen:14,xsl-gonneau2020,Verro:22}. The current version of the PAUS calibration, uses the second data release (DR2) of the XSL library containing 666 templates, a subsample of which is shown in Fig.~\ref{fig:xsl_seds}. 

\begin{figure}
  \centering
  \includegraphics[width=\linewidth]{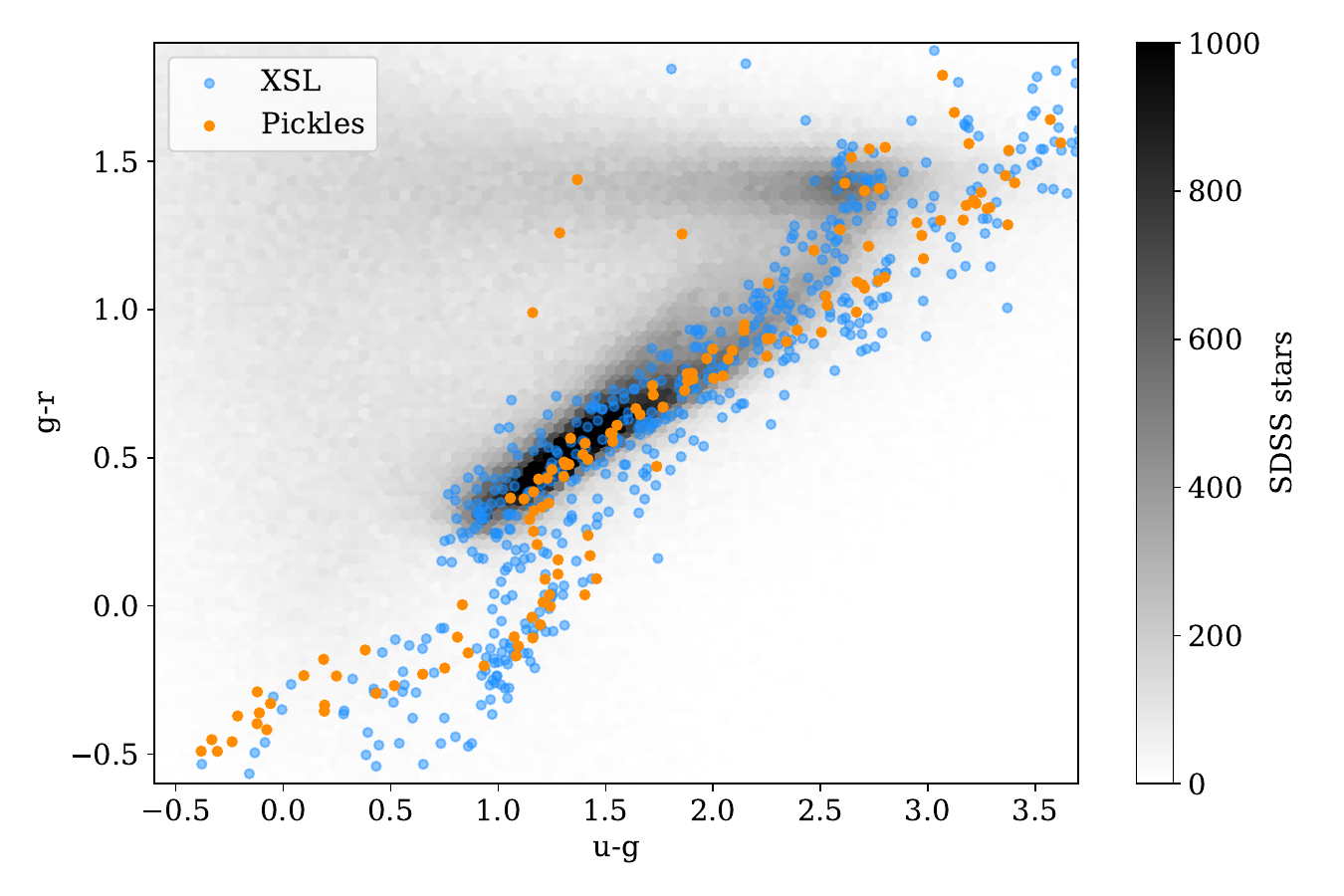}
  \caption{The (g-r).vs.(u-g) colour-colour plot of the SDSS stars (grey distribution) with  the XSL DR2 stellar templates (blue points). For comparison, we also show the widely used Pickles stellar library~\citep{stars-pickles1998} as orange points.  An average Galactic extinction was added to the templates to allow a direct comparison with the extincted magnitudes of SDSS.}
  \label{fig:sdss_pickles_xsl}
\end{figure}

\begin{figure}
  \centering
  \includegraphics[width=\linewidth]{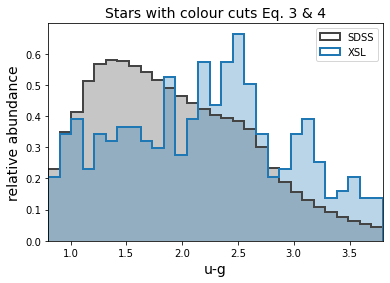}
    \includegraphics[width=\linewidth]{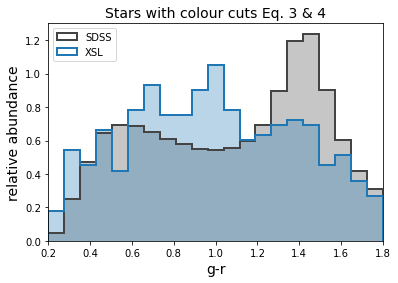}
  \caption{Normalised colour histograms of the SDSS stars (grey distributions) and the XSL DR2 stellar templates (blue distributions) plotted in Fig.~\ref{fig:sdss_pickles_xsl} applying the cuts in equations~\ref{eq:colour_cut_qso1} and~\ref{eq:colour_cut_qso2}. The top panel is for the $(u-g)$ colour and the bottom panel for the $(g-r)$ colour. Both distributions are similar. The Pickles stellar library is omitted for clarity. }
  \label{fig:sdss_xsl_hist}
\end{figure}

The XSL stellar library contains mid-to-high resolution ($R\sim10\,000$) spectra from the near-ultraviolet to the near-infrared of a comprehensive set of stars. The synthesized colours from the spectra compare well with those obtained from Gaia and other stellar libraries~\citep{xsl-gonneau2020,Verro:22}. The library was created to cover a wide range of spectral types and chemical compositions. The Fig.~1 of~\cite{Verro:22} shows the wide range of stellar atmospheric parameters covered by the library. Figure~\ref{fig:sdss_pickles_xsl} shows the XSL sampling of the colour-colour stellar locus compared to the SDSS DR12 stellar sample. The main differences are the absence of very blue stars (e.g., $g-r < 0.3$) in the SDSS and the excess of stars blue in ($u-g$) and red in $(g-r)$ in the SDSS with respect to the XSL library. These former early type stars have already evolved away from the main sequence in non-star forming regions in our galaxy. They are, anyway, excluded from our sample to avoid confusion with QSOs (Eqs.~\ref{eq:colour_cut_qso1} and~\ref{eq:colour_cut_qso2}).  The later stars are normally due to unresolved binaries in the SDSS catalogue. Figure~\ref{fig:sdss_xsl_hist} shows the colour histograms of the SDSS and XSL stars shown in Fig.~\ref{fig:sdss_pickles_xsl}. The histograms are somewhat different as the SDSS sample is a representation of the density of stars in high galactic latitude fields while the XSL was constructed to sample the Hetzsprung-Russell (HR) diagram more evenly. For our calibration purposes, the important point is that the SDSS stars we use as calibrators can be properly modeled as a linear combination of the XSL stellar library.

\subsection{Synthetic PAUS magnitudes}
\label{s:synth_pau_mags}

Next, we compute the synthetic PAUS 40 NB magnitudes for the XSL stellar template set. In order to do that, we need detailed knowledge of the transmission of the PAUCam system, including the primary mirror reflection, the transmission of the optical corrector and our filters and the response of the CCDs. We used the values obtained in the calibration of the PAUCam instrument and presented in \citet{Padilla:2019}. 

To improve efficiency, we pre-compute the synthetic fluxes from all the XSL templates for all the PAUS narrow band filter transmissions and also the SDSS filters. This way, we can fit the stellar template that best fits the SDSS photometry and then obtain the PAUS magnitudes for that template, using the pre-computed values instead of performing the integrals for every star at each computation step.

In order to obtain a robust calibration, we use only stars that are detected in the PAUS images at a signal-to-noise S/N $> 10$ and with SDSS $i_{AB} < 21$. We also avoid stars that are saturated. The typical magnitude range that we use is $14<i_{AB}<21$. In order to estimate whether the Milky Way dust extinction should be taken into account, we note that main sequence stars from F to K-type have a range of absolute magnitudes $M_i\sim 2.5-6.0$. 
These stars will be observed with magnitudes fainter than $i\sim16$, if located 1 kpc away from us. 
So, most of the SDSS stars we use for calibration are located at distances larger than 1 kpc and are attenuated by Milky Way extinction. The XSL DR2 template stars are not corrected for extinction. However,
they have magnitudes that are typically $\sim$10 magnitudes brighter than the SDSS stars, which corresponds to being 100 times closer in luminosity distance if they have the same luminosity. Therefore, they experience less extinction and we need to take this difference into account.
Consequently, we also compute the synthetic fluxes of the XSL templates at a range of extinction values from the lowest to the largest value in each field in steps of 0.005 in colour excess $E(B-V)$ following the O'Donnell extinction model \citep{extinction-odonnell1994} assuming $R_{\rm v} = 3.1$. 

As the template units are defined in flux density $f_\lambda$ with units of erg cm$^{-2}$ s$^{-1}$ \AA$^{-1}$ we can compute the integrated average photon flux density of the template at the specific passband as:
\begin{equation}
F(t, R_i, A_{\rm v}) = \frac{\int f_{\lambda_t} R_i(\lambda) A(\lambda) \lambda^2 d\lambda}{c \int R_i(\lambda) d\lambda}  \;,
\end{equation}
where $\lambda$ is the wavelength, $t$ denotes the stellar template, $c$ is the speed of light, $R_i$ is the filter transmission of reference\footnote{Note that the definition used here for $R_i(\lambda)$ differs from the one used in~\cite{Fukugita:96} for $R_{\nu}$ in the definition of the AB system for filters. They are related as $R_{\nu} = R_i(\lambda)\, \lambda$. This same definition is used in section~\ref{s:synthetic_photometry}.} and $A_{\rm v}$ is the extinction normalisation applied to the O'Donnell extinction law to obtain the extinction $A(\lambda$) in units of flux as a function of wavelength.  

For the XSL DR2 set of 666 stellar templates, 40 PAU narrow bands + 5 SDSS broad bands and typically 10 extinction values, we end up with $\sim300\,000$ pre-computed synthetic template fluxes.

\subsection{Source extraction}

\begin{figure}
  \centering
  \includegraphics[width=\linewidth]{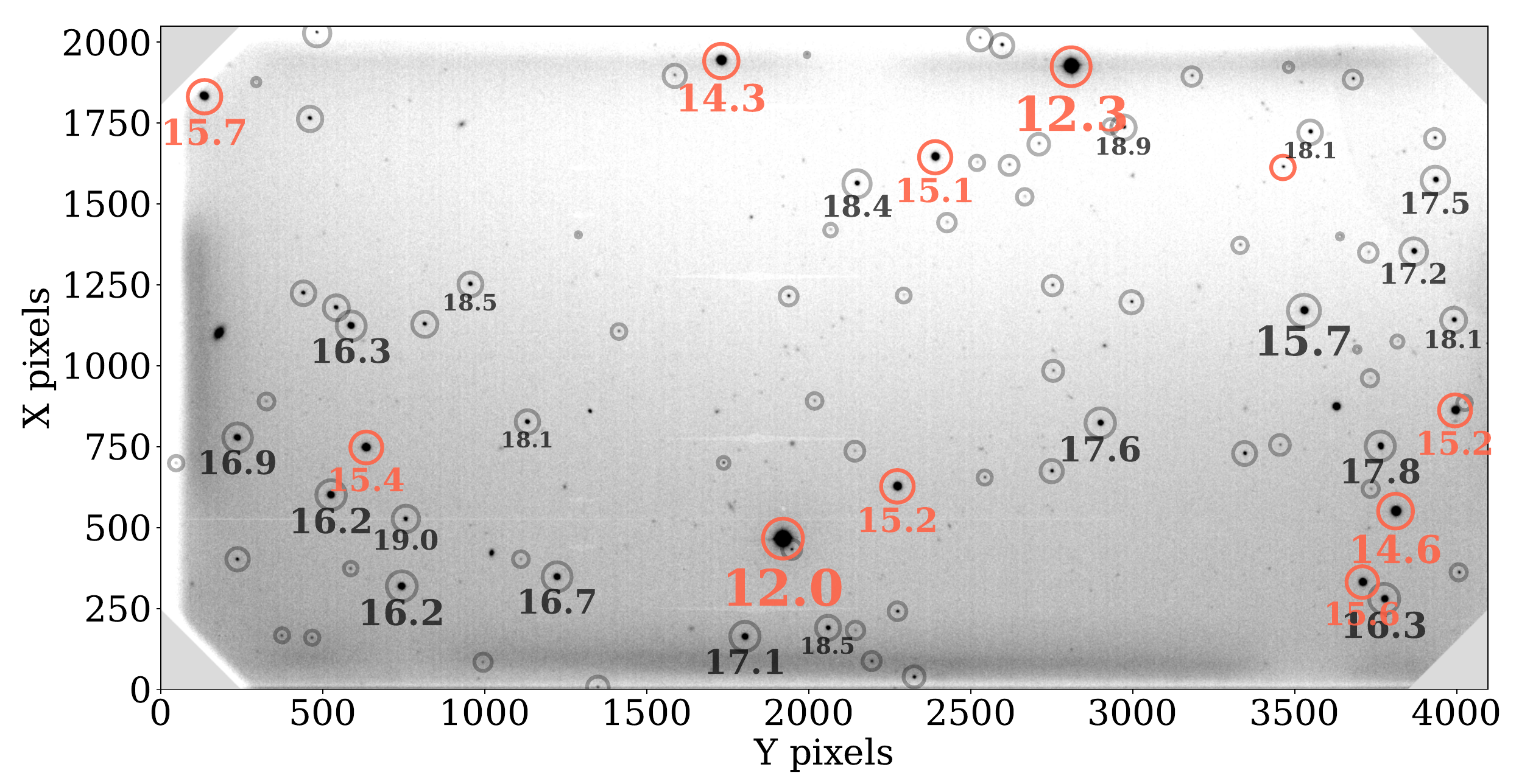}
  \caption{An image in the  NB565 filter with stars that were matched between PAUS and SDSS. Y pixels running horizontally and X pixels running vertically in the CCD axis convention. The numbers in the image correspond to the $i_{AB}$ magnitude measured by SDSS. The distribution ranges from very bright stars of 12th magnitude to magnitude 21. Grey circles indicates ``clean" SDSS photometry while red ones indicate that the photometry is marked as ``not clean". Most very bright stars are marked as "not clean" and are not used in the calculation of the photometric zero points.}
  \label{fig:sdss_calib_stars_image}
\end{figure}

To obtain the uncalibrated fluxes of the stars measured in the PAUS narrow bands, we run \texttt{Source Extractor}~\citep{sextractor-bertin1996} over the instrumentally detrended and astrometrically calibrated images~\citep{Serrano:2022}. Our target calibration stars are moderately bright, comprising magnitudes between 14 and 21, that typically are observed with a signal-to-noise ratio higher than 10. For such bright stars we do not need to optimize the aperture with complex and PSF-dependent methods that could be sensitive to the observing conditions or optical distortions in the focal plane. Instead we aim for a constant large aperture ($\sim$4" radius) that gathers almost all of the light from the star independently of its PSF, ensuring that more than 99\% of the flux is within the aperture in the worst seeing conditions tolerated by the survey. We tested various configurations of aperture sizes, background modelling and scattered light correction. The method described previously proved to be the most reliable across the different observing conditions. Once the photometry is processed, we perform a spatial matching with the SDSS DR12 \citep{sdssdr12-2015} catalogue, as only those stars will be of interest for the photometric calibration of the narrow band images. An example of a narrow band image of PAU with the calibration stars from SDSS marked can be found in Fig.~\ref{fig:sdss_calib_stars_image}.

\subsection{Statistical methodology}
\label{subsec:statistical_methodology}

The next step is to fit the stellar templates to the five $ugriz$ bands SDSS star PSF fluxes for each star, taking into account their associated errors. 
From the sky coordinates of the star we can obtain its Galactic extinction value from our reference reddening dust map \citep{dust-planck2014} and use the pre-computed synthetic template SDSS fluxes at that particular extinction to compute the 
$\chi^2$ goodness of fit between templates and observed data as
\begin{equation}
\chi^2_{\rm t} = \sum_{\rm b}  \frac{(f_{\rm obs}^{\rm b} - \alpha_{\rm t} f_{\rm t}^{\rm b})^2}{({\sigma_{\rm obs}^{\rm b}})^2} \;,
\end{equation}
where the summation is over the five SDSS bands denoted by index b, $f_{\rm obs}^{\rm b}$ are the observed fluxes in each band b, with $\sigma_{\rm obs}^{\rm b}$ being their errors, $f_{\rm t}^{\rm b}$ are the synthetic template fluxes including extinction for each band b, and $\alpha_{\rm t}$ is the scaling factor that minimizes the  $\chi^2_{\rm t}$ for template t that is computed as
\begin{equation}
\alpha_{\rm t} = \frac{\sum_{\rm b} f_{\rm obs}^{\rm b} f_{\rm t}^{\rm b} \frac{1}{({\sigma_{\rm obs}^{\rm b}})^2 }}{\sum_{\rm b} ({f_{\rm t}^{\rm b}})^2 \frac{1}{({\sigma_{\rm obs}^{\rm b}})^2 }}  \;,
\end{equation}
and its associated error $\sigma_{\alpha_{\rm t}}$
\begin{equation}
\sigma_{\alpha_{\rm t}} = \frac{1}{\sqrt{ \sum_{\rm b} \frac{f_{\rm t}^{\rm b}}{({\sigma_{\rm obs}^{\rm b})}^2}}} \;.
\end{equation}

We could estimate the reference narrow band flux of a particular star, $f_{\rm star}$, using the best-matching spectral energy distribution, $\alpha_{\rm best} f_{\rm best}$. 
However, selecting the "best" $\chi^2$-matching star can lead to systematic errors in the calibration, as the template basis does not necessarily sample the whole parameter space of spectral energy distributions of the stars observed by the SDSS. As an indication, Fig.~\ref{fig:sdss_pickles_xsl} shows the distribution of the XSL DR2 star templates in the $u-g$ vs $g-r$ colour space on top of the observed colours of the SDSS stars.
Therefore, we take into account all the templates weighted by their goodness of fit. 

The probability of a template fit can be defined as
\begin{equation}
P_t = \frac{e^{-\chi_{\rm t}^{2}/2}}{ \sum_{\rm t} e^{-\chi_{\rm t}^{2}/2 }} \;,
\end{equation}
and the estimate of the expected flux at a given band b is then
\begin{equation}
\langle f_{\rm star}^{\rm b} \rangle = \sum_{\rm t} P_{\rm t} \alpha_{\rm t} f_{\rm t}^{\rm b} \;,
\end{equation}
and the estimated flux error 
\begin{equation}
\sigma_{\langle f_{\rm star}^{\rm b} \rangle} = \sum_{\rm t} P_{\rm t} \sigma_{\alpha_{\rm t}} f_{\rm t}^{\rm b} 
\end{equation}

This estimated flux takes into account all templates and fits. It is more robust to uncertain situations where not a clear best fit is defined. The observed flux in band b, $f_{\rm obs}^{\rm b}$, of each star in a PAUS narrow band image that matches an SDSS star provides a calibration zero point for that band that can be defined as
\begin{equation}
\text{ZP}_{\rm star}^{\rm b} = \frac{\langle f_{\rm star}^{\rm b} \rangle }{f_{\rm obs}^{\rm b}} \;,
\label{eq:ZP_star_b}
\end{equation}
and its associated error
\begin{equation}
\sigma_{\text{ZP}_{\rm star}^{\rm b} }= \frac{\sigma_{\langle f_{\rm star}^{\rm b} \rangle}}{f_{\rm obs}^{\rm b}} \;.
\end{equation}

The compute image zero point in band b as the mean average of the star zero points in the image

\begin{equation}
\text{ZP}_{\rm image}^{\rm b} = \frac{1}{N}\sum_{\rm star} \text{ZP}_{\rm star}^{\rm b}  \;,
\label{eq:zpimage}
\end{equation}
where $N$ is the number of ZP$_{\rm star}^{\rm b}$ values used in calculating the mean. Typically, there are 50-80 stars contributing to the sum depending on the band, observing conditions and sky coordinates. We have chosen to use an unweighted mean because the weighted sum was providing worse results. We believe this is due to uncontrolled systematic errors in the measurement of the brightest stars.

We estimate the image zero point error calculating 
the mean zero point for 200 bootstrap realizations ($\rm ZP_{boot}^{\rm b}$) of the
star zero points in each image corresponding to band b. The image zero point error
is then estimated as the difference
\begin{equation}
\label{eq_ZPerror}
\sigma_{\text{ZP}_{\rm image}^{\rm b}} = 0.5 \left[\left(\text{ZP}^{{\rm b}}_{\rm boot}\right)^{84} - \left(\text{ZP}^{\rm b}_{\rm boot}\right)^{16} \right] \;,
\end{equation}

\noindent
of the 84\% and 16\% quantiles of the corresponding distribution of the zero-points estimated by the bootstrap realizations ($\text{ZP}_{\rm boot}^{\rm b}$). This method reduces the dominance of outliers and 
 avoids relying on propagating an analytical error estimate,
 which is susceptible to underestimated errors.

This process will return a star zero point, $\text{ZP}_{\rm star}^{\rm b}$, the zero point error, $\sigma_{\text{ZP}_{\rm star}^{\rm b}}$, and the best template $\chi^2_{\rm best}$ fit for each star in the image, as well as the combined image zero point in band b, $\text{ZP}_{\rm image}^{\rm b}$ and its estimated zero point error, $\sigma_{\text{ZP}_{\rm image}^{\rm b}}$. All values are stored in the PAU database for later processing and analysis. As the PAU data management works with fluxes in all its measurements, the zero points are all defined as a multiplicative factor in flux rather than an additive magnitude.

The calibrated single-epoch flux for each object in band b is then estimated using the image zero point calibration as
\begin{equation}
\label{eq:f_calib}
f_{\rm calib}^{\rm b} = f_{\rm obs}^{\rm b} \,  \text{ZP}_{\rm image}^{\rm b}
\end{equation}
and its calibrated error, $\sigma_{f_{\rm calib}^{\rm b}}$, assuming non-linear error propagation with independent and non-negligible variances, is derived as
\begin{equation}
\sigma_{f_{\rm calib}^{\rm b}} = \sqrt{(\sigma_{\rm obs}^{\rm b})^2  (\sigma_{\text{ZP}_{\rm image}^{\rm b}})^2 + (\sigma_{\rm obs}^{\rm b})^2  (\text{ZP}_{\rm image}^{\rm b})^2 + (f_{\rm obs}^{\rm b})^2(\sigma_{\rm obs}^{\rm b})^2} 
\label{eq:caliberror}
\end{equation}
where $f_{\rm obs}^{\rm b}$ and $\sigma_{\rm obs}^{\rm b}$ are the observed fluxes and standard deviations, respectively, and $\text{ZP}_{\rm image}^{\rm b}$ and $\sigma_{\text{ZP}_{\rm image}^{\rm b}}$, the zero point calibration (Eq.~\ref{eq:zpimage}) and its error (Eq.~\ref{eq_ZPerror}). Equation~\ref{eq:caliberror} neglects the covariance between the observed fluxes and the image zero point determination. This term is negligible in most cases, except maybe for the brightest stars.  


\section{Validation and performance}
\label{section: Validation and performance}

The scientific performance of the PAU Survey relies on how well we can calibrate the data. To estimate how well the calibration technique performs, we consider several tests

\begin{enumerate}
\item Check the dependence of the zero points with respect to airmass (Sec.~\ref{subsec:image_ZP_airmass})
\item Compare the individual star zero points to the averaged image zero points (Sec.~\ref{subsec:star_image_ZP}).
\item Compare our calibrated fluxes to synthetic narrow band photometry from observed spectra that overlap with our images (Sec~\ref{subsec:comparison_SDSS_Spectra}).
\item Compare duplicate measurements (Sec.~\ref{subsec:duplicates}).
\item Compare with the calibration obtained from spectrophotometric standard stars observed during the survey (Sec.~\ref{subsec:spectrophotometric}).
\end{enumerate}

\subsection{Image zero point vs airmass}
\label{subsec:image_ZP_airmass}
In section~\ref{subsec:general_procedure}, we mentioned that in a photometric night the zero point (in magnitudes) depends linearly on airmass with a slope given by the extinction coefficient (equation~\ref{eq_ZP}). The image zero point, $\text{ZP}_{\rm image}^{\rm b}$ in equation~\ref{eq:zpimage}, that we compute following the procedure in section~\ref{subsec:statistical_methodology} corresponds to the zero point, ${\rm ZP}_1$ in equation~\ref{eq_ZP}. 

\begin{figure}
  \centering
  \includegraphics[width=1.\linewidth]{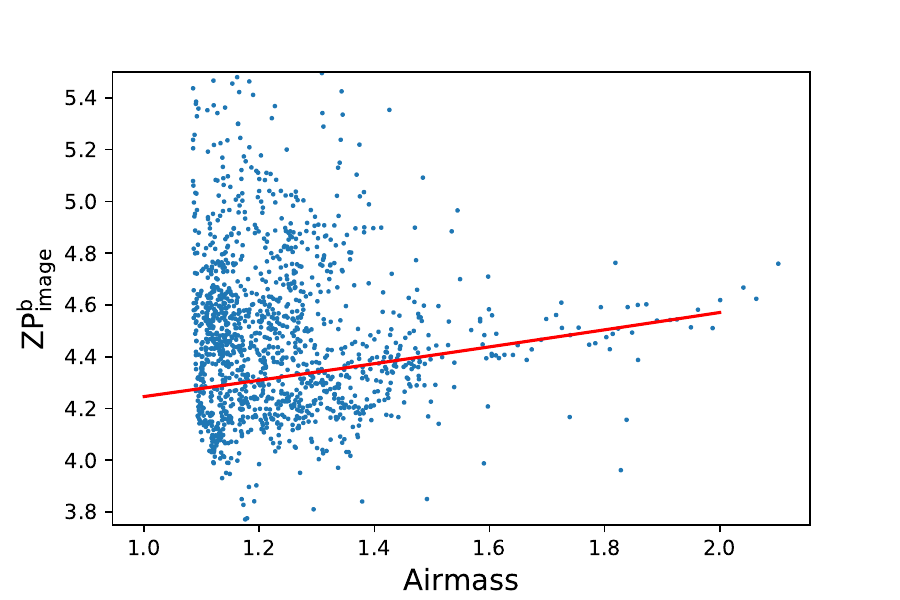}
  \caption{Image zero points, $\text{ZP}_{\rm image}^{\rm b}$, as a function of airmass for observations taken with the filter NB615 as blue dots. In red we show the expected behaviour of the zero point with airmass following the extinction coefficient for that wavelength given in King (1985) (footnote~\ref{footnote:technote}) with an arbitrary normalisation.}
  \label{fig:ZP_mult_vs_airmass}
\end{figure}

Figure~\ref{fig:ZP_mult_vs_airmass} presents the image zero point, $\text{ZP}_{\rm image}^{\rm b}$, as a function of airmass for observations taken with the filter NB615. We also show for comparison as a red line the expected dependence of the zero point with airmass following the extinction coefficient for that wavelength provided by King (1985)\footnote{La Palma Technical Note \#31  \url{http://www.ing.iac.es/Astronomy/observing/manuals/ps/tech\_notes/tn031.pdf}\label{footnote:technote}}.
The PAUS observations have been taken in a wide variety of observational conditions. Those variations produce the wide spread in the image zero points at the same airmass. Most of the variation is probably due to high cirrus that induce wavelength independent opacity and significant spatial structure. An analysis of the spatial structure function supports this interpretation.
Although difficult to estimate from the wide dispersion of zero points, the ZPs not affected by extinction seem to 
be consistent with the expected behaviour as a function of airmass.

\subsection{Star vs image zero point}
\label{subsec:star_image_ZP}

We calibrate the data with the combined image zero point constructed from the individual star zero points. 
The star ZP measurements can be individually inaccurate but given the large number of measurements provide valuable information that allow us to perform both spatial and wavelength corrections over the baseline image zero points.

\subsubsection{Background subtraction effects}

\begin{figure}
  \centering
  \includegraphics[width=1.\linewidth]{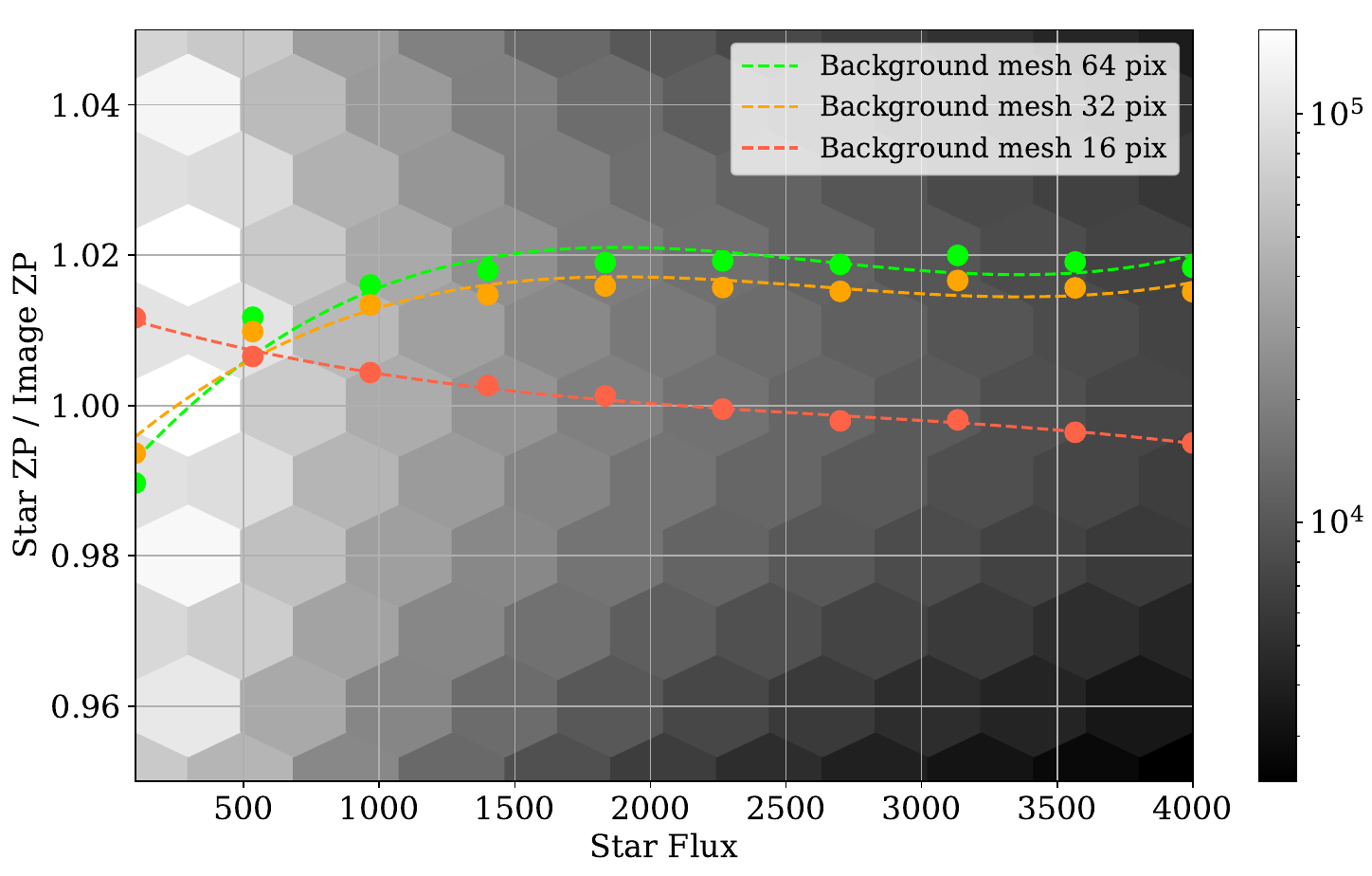}
  \caption{Ratio of the individual star ZP, $\text{ZP}_{\rm star}^{\rm b}$, to the corresponding image ZP, $\text{ZP}_{\rm image}^{\rm b}$, plotted as a function of the star flux in ${\rm e}^{-}/s$ for three values of the background mesh (64, 32 and 16 pixels). A 3rd order polynomial fit is adjusted to the average values to identify the non-linear trend at each configuration. The gray scale indicates the number of star ZPs per hexagon.}
  \label{fig:star_image_zp_flux}
\end{figure}

The PAUS images suffer from considerable contamination from scattered light at the edges of each CCD~\citep{Padilla:2019,Serrano:2022}. In order to check for possible biases due to our background estimation, 
we ran the photometry of stars with three different background subtraction scales to correct for scattered light residuals. By default \texttt{SExtractor} uses a mesh of 64 pixels to account for large variations in the background. Scattered light produces variations at smaller scales than regular background variations and therefore we need to set a smaller value of the mesh. However, setting a value that is too small subtracts flux from the source instead of the background, altering the measurements and therefore delivering less accurate photometric zero points. 
In Fig.~\ref{fig:star_image_zp_flux} we show the ratio of the star ZP, $\text{ZP}_{\rm star}^{\rm b}$ to the image ZP, $\text{ZP}_{\rm image}^{\rm b}$, as a function of star flux for three different background subtraction mesh sizes. As seen in Fig.~\ref{fig:star_image_zp_flux}, a mesh size of 16 pixels delivers a ratio of the individual calibration stars ZP to the overall image ZP image closer to 1 across different flux ranges. This is only valid for point-like sources, as extended galaxies may be too large for such a small background mesh and therefore we increased the mesh size to 32 for all galaxy photometry obtained with our Multi-Epoch Multi-Band Aperture photometric pipeline~\citep[\textsc{MEMBA},][]{Serrano:2022}, which combines the single image measurements.

At the scale used for the calibration of the star photometry (16 pixels), there is a non-linear relation with the star flux, resulting in a 2\% variation from the low to mid range star flux (Fig.~\ref{fig:star_image_zp_flux}). Even though this strong background subtraction is moderately biasing the estimated ZP of the brightest stars, the flatter scatter light-free image with this sky subtraction delivers individual ZPs that do not depend as much on the overall stellar flux compared to smoother background subtractions.

\subsubsection{Effects due to stellar types}
\label{s:blue_stars}

\begin{figure}
  \centering
  \includegraphics[width=1.\linewidth]{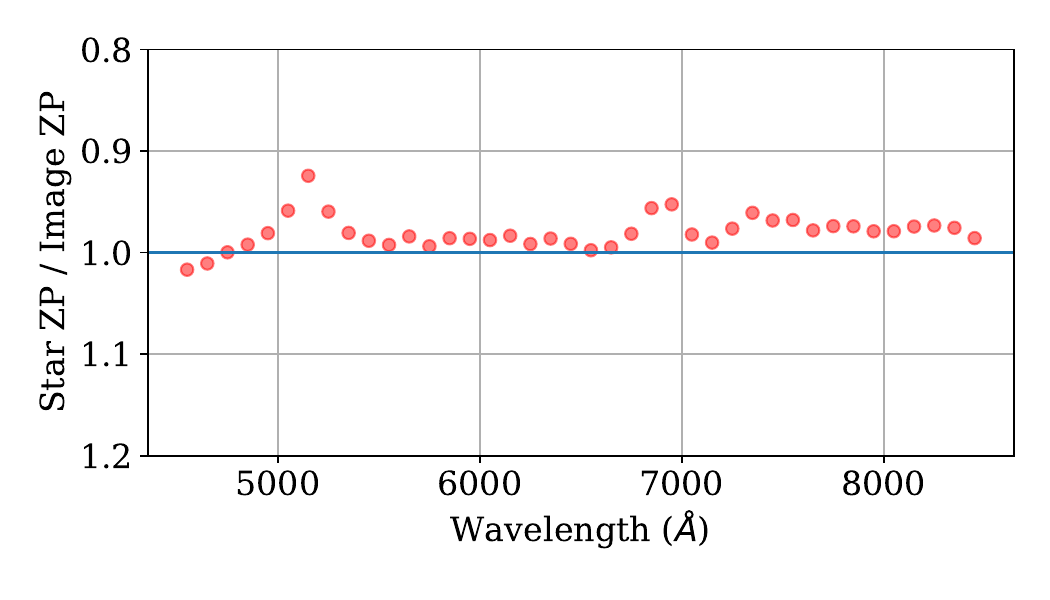}
  \caption{Ratio of the image ZP, $\text{ZP}_{\rm image}^{\rm b}$ computed with only blue stars to the image ZP, $\text{ZP}_{\rm image}^{\rm b}$, computed  with all the stars for a subsample of blue stars selected with  $g-r < 0.4$ and in the flux range of 2000 to 5000 (${\rm e}^-/s$). Blue stars have smoother spectra than red stars, and therefore they provide a better global ZP correction as they do not have the large short wavelength variations that redder stars present for the same broad band colours.}
  \label{fig:star_image_zp_blue}
\end{figure}

Bluer hotter stars have smoother spectra with fewer absorption and emission features than redder cooler stars. Their high frequency SED variations for the same broad band colours are smaller providing a better calibration reference for the narrow band system. However, we cannot rely solely on blue stars as this would reduce the reference catalogue to a small fraction of the full sample, leaving some individual images with too few or even no calibration stars. In order to test the effects of SED variations for similar broad band colours, we select a subsample of blue stars by applying a $g-r < 0.4$ colour cut (note that we have already applied a colour cut to reject QSOs that also rejects the bluest stars, Eq.~\ref{eq:colour_cut_qso1} \&~\ref{eq:colour_cut_qso2} and Fig.~\ref{fig:qso_rejection}), which removes 90\% of the whole sample, and compute the zero point for this subsample of stars with smoother SEDs that make the estimation of the narrow band fluxes more robust. We obtain the ZP of the image computed with only blue stars and compare it to the image ZP obtained with all the stars in the flux range of 2000-5000 electrons/s, where this ratio is flatter and does not have a strong dependency on the background subtraction technique, as explained in the previous subsection (see Fig.~\ref{fig:star_image_zp_flux}). The offsets found for each narrow band are shown in Fig.~\ref{fig:star_image_zp_blue}, enabling global corrections to the individual image zero points of each band. The bump around 5170 {\AA} is expected to be due to the MgI absorption triplet present in the redder stars that depends mainly on the star metallicity. The same broad-band colours can have different strengths of this absorption feature and produce different narrow-band fluxes at these wavelengths, resulting in an uncertain calibration.
The redder filters, with wavelengths larger than 6800 {\AA}, are affected by telluric absorption and emission lines from the atmosphere. These features are variable and can influence the calibration at those wavelengths. The XSL spectra are corrected for sky emission and telluric absorption~\citep{xsl-gonneau2020} and in principle should not cause any calibration systematic feature. Nevertheless, the fitting procedure calibrating from broad band photometry to templates can have larger variations for the reddest stars and may be responsible for the deviations from equality seen in the comparison of the calibration performed with the bluer stars to the calibration with all the stars shown in Fig.~\ref{fig:star_image_zp_blue}.

\subsubsection{Spatial variation effects}

\begin{figure}
  \centering
  \includegraphics[width=1.\linewidth]{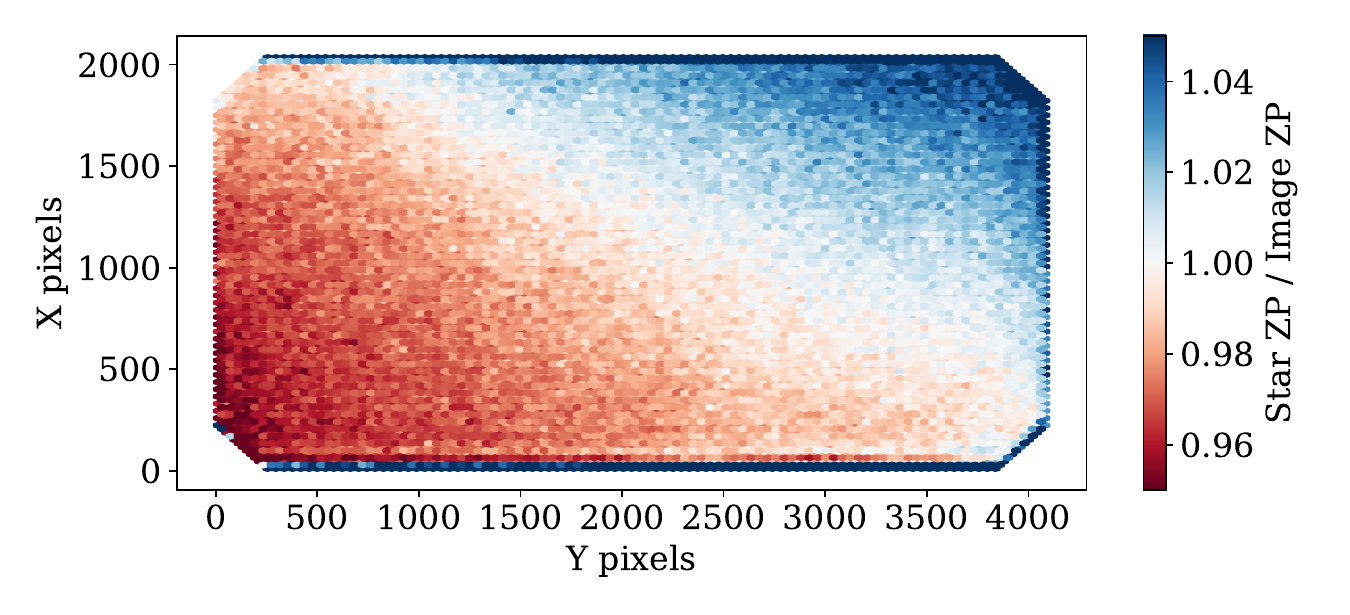}
  \caption{The star ZP, $\text{ZP}_{\rm star}^{\rm b}$, to the 
 total image ZP, $\text{ZP}_{\rm image}^{\rm b}$, ratio in the CCD detector 1 of PAUCam. The colour code shows the value of this ratio. The narrow bands that correspond to this detector (NB455, NB535, NB685, NB765 and NB845) show the same pattern across the detector, which suggests that this effect is due to an illumination pattern that does not depend on wavelength.}
  \label{fig:star_image_zp_spatial}
\end{figure}

Finally, we evaluate the spatial dependence of the star-to-image ZP ratio across the field of view. The large field of view of PAUCam suffers from strong optical vignetting caused by the WHT corrector lens, which reduces the detected amount of light at the edge of the eight central detectors, where the narrow band filters are installed~\citep{Padilla:2019}, and almost fully at the edge of the 18 detector mosaic array. We use the dome flat fields to correct for the illumination pattern of the focal plane. In Fig.~\ref{fig:star_image_zp_spatial} we show the residuals of the individual star ZPs to the global image ZP for one of the CCDs at the edge of the eight central detectors where the NB filters are located. We can observed an 8\% variation from one side to another in the detector. We can use this measured variation to build a star flat correction to account for the large scale illumination pattern and correct for it at the catalogue level.

\subsection{Comparison with SDSS spectra}
\label{subsec:comparison_SDSS_Spectra}

In the PAUS narrow band (NB) calibration, we have used the broad band (BB) photometry from the SDSS imaging survey. This means that the PAUS NB calibration is tied to the SDSS photometric system.
Here, we will check how the calibration may change if we use the SDSS spectra instead. This is not a redundant or trivial test. The SDSS imaging is taken with a different instrument than the SDSS spectra albeit using the same telescope and site. Even though there is a cross-calibration between the two SDSS modes of operation, the noise, systematics, observing conditions, apertures and wavelength resolution are different. 
We will use the higher spectral resolution to validate whether the PAUS calibration method that we have designed is uniform as a function of wavelength and whether there are high-resolution wavelength features not taken into account. The uniformity of the wavelength calibration is a key feature for the PAUS survey as we need well-calibrated SEDs for all objects across the sample to reach our goals.

\begin{figure}
  \centering
  \includegraphics[width=.8\linewidth]{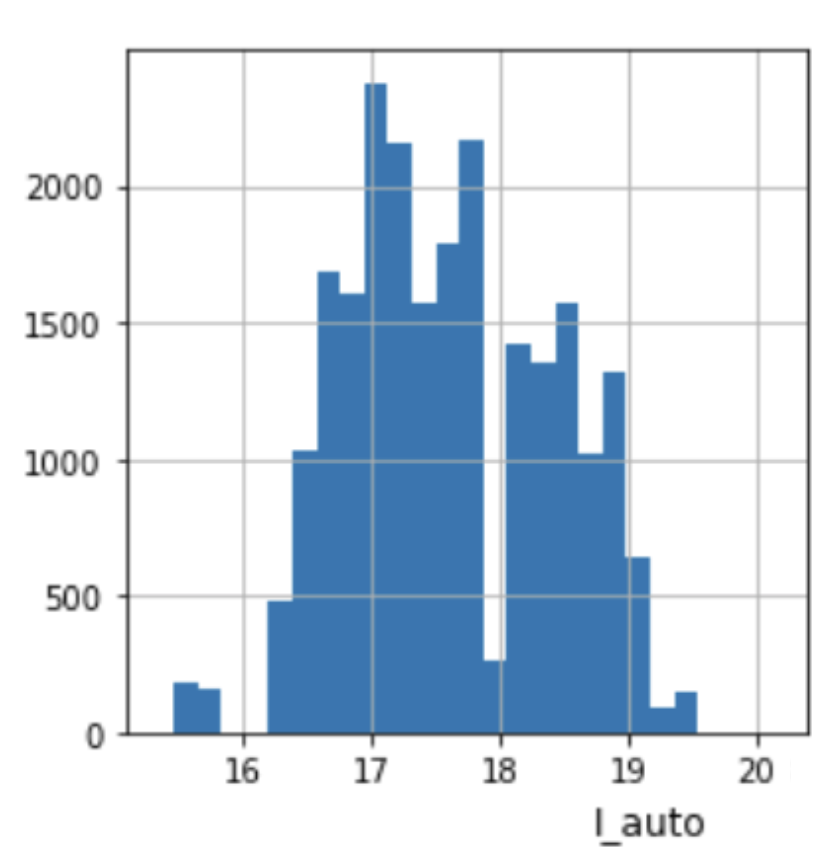}
  \caption{The distribution of i-band ($I_{\text{auto}}$ in COSMOS) values for stars in the COSMOS field used for the validation.}
  \label{fig:NIauto}
\end{figure}

The SDSS programme has run in several stages conducting large optical spectroscopic programmes. Luckily, there is significant overlap of the PAUS wide fields with their spectroscopy. 
We selected a sample matching the PAUS calibration stars with the SDSS spectra which are also in the COSMOS reference catalogue~\citep{Laigle:16} for validation.
Figure~\ref{fig:NIauto} shows a histogram of the total $i$-band magnitude\footnote{$I_{\text{auto}}$ in the COSMOS reference catalogue} for the $\sim$200 stars that we use in our validation. 

\begin{figure}
  \centering
  \includegraphics[width=1.\linewidth]{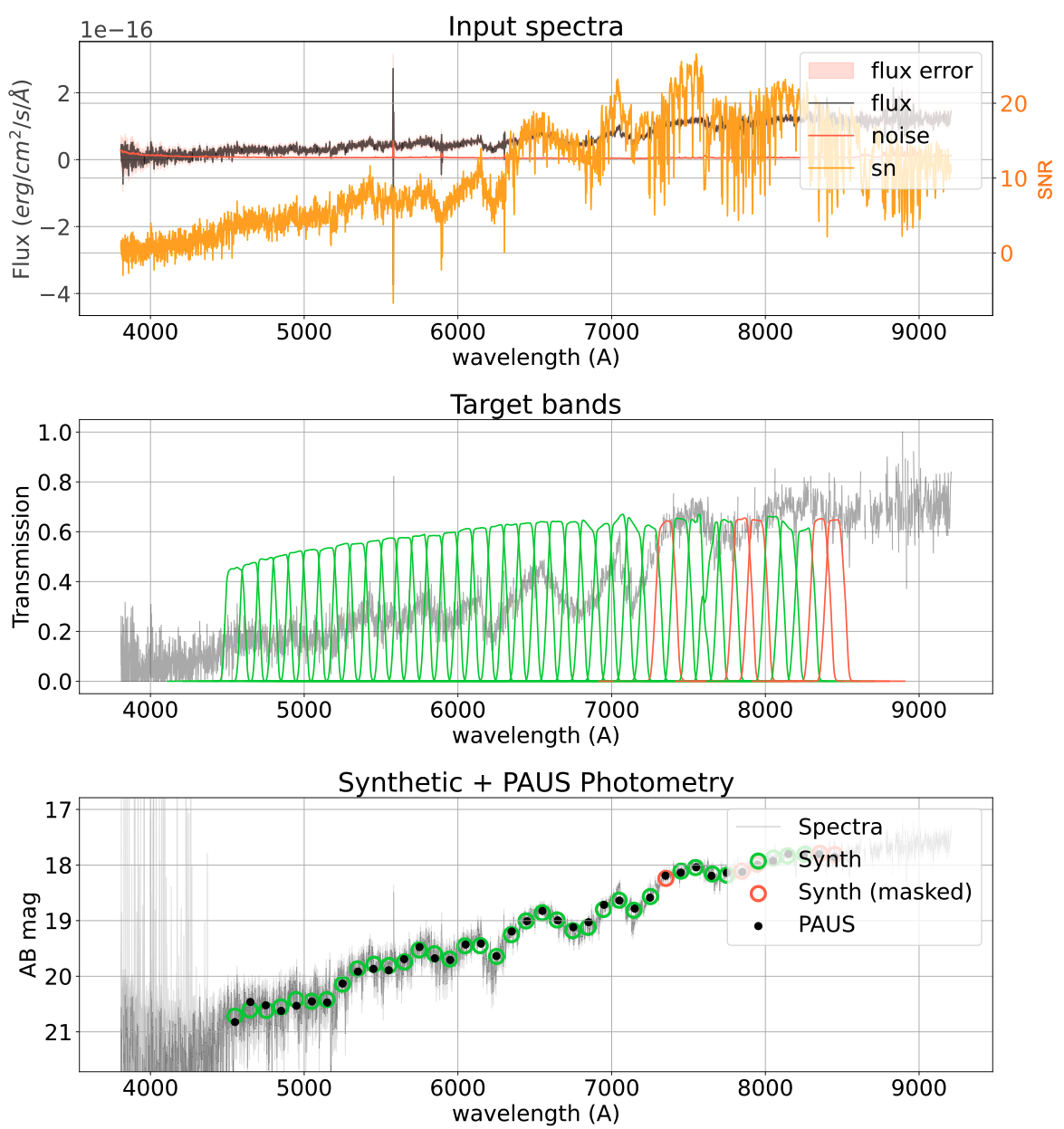}
  \caption{Top panel: an example of an M star spectrum from SDSS used as input for synthetic photometry. The black line is the flux density spectrum in units of erg/cm$^2$/a/{\AA}. The red line shows the noise and the orange line the signal-to-noise. Middle panel: Stellar spectrum in gray with the transmission of the 40 PAU narrow band filter set in green and red. Finally, the bottom panel shows again stellar spectrum in gray and the computed synthetic photometry from the spectrum as green open circles. The bands without enough unmasked samples from the spectrum are marked as red open circles. The black dots are the observed PAUS measurements.}
  \label{fig:sdss_input_spectra}
\end{figure}
\subsubsection{Synthetic narrow band photometry}
\label{s:synthetic_photometry}

Similarly to the process of obtaining synthetic PAUS narrow band fluxes from the SDSS broad band photometry using the stellar templates (detailed in \S\ref{s:synth_pau_mags}), we compute the synthetic narrow band photometry in the PAU filters from SDSS spectra. The process is illustrated in Fig.\ref{fig:sdss_input_spectra}.

The initial step in the process of generating the synthetic bands consists on retrieving and homogenizing the spectral data. In our case all flux measurements are converted to a common flux density, $f_\lambda$, in units of erg/cm$^2$/s/{\AA}.
Generally, each spectrum contains the coverage wavelength range, the flux, the noise (or inverse variance) and a mask. Optionally, SDSS also includes a measurement of the sky spectrum, that allows the user to identify possible contamination by strong emission or absorption lines. In a second stage, we interpolate the filter transmission response, $R(\lambda)$, to the wavelength sampling of the spectral data.
We mask the wavelengths 
that are flagged in the spectral mask and compute the integrated average flux density of the source at the specific passband filter in erg/cm$^2$/s/Hz such as:

\begin{equation}
\label{eq:fnu_mean}
\langle F_\nu \rangle = \int \frac{f_\lambda R(\lambda) \lambda^2}{c}  d\lambda \;,
\end{equation}
and its associated integrated response
\begin{equation}
\label{eq:response_filter}
R_i = \int R(\lambda) d\lambda \;.
\end{equation}

Finally, we compute the synthetic magnitudes in the AB system with the following transformation:

\begin{equation}
\label{eq:AB_filter}
m_{\rm syn} = -2.5 \log_{10} \left(\frac{\langle F_\nu \rangle}{ R_i }\right)  -48.6 \;.
\end{equation}

Note that the response filter $R(\lambda)$ in equations~\ref{eq:fnu_mean}, \ref{eq:response_filter} and \ref{eq:AB_filter} is not the same as the response filter $R_{\nu}$ in the standard AB system definition of~\cite{Fukugita:96}. See section~\ref{s:synth_pau_mags}.

It is also important for the statistical analysis to estimate the error of each synthetic photometric estimate. As the flux in the spectra has been weighted by the response of the transmission, we must weight the noise in the spectra by the relative transmission throughout for the entire passband.

\begin{equation}
\sigma_{f_\nu}^2 = \int  \frac{R(\lambda)^2 \sigma_\lambda^2 \lambda^2 }{c^2 R_i^2}  d\lambda \;,
\end{equation}
where $\sigma_\lambda$ is the noise in the high-resolution spectra. 
We can approximate the magnitude error as

\begin{equation}
\sigma_{m_{\rm syn}} \approx 2.5\,\log_{10}({\rm e})\,  \frac{\sigma_{f_\nu}^2}{\langle F_\nu \rangle/ R_i } \;.
\end{equation}

Following the previous procedure, where the syntehic magnitude $m_{\rm syn}$ is the equivalent to the template flux $f_{\rm t}$ of section~\ref{subsec:statistical_methodology}, we compute the synthetic photometry over all the SDSS stellar spectra that overlap with PAUS over the 40 PAUS narrow band set and the SDSS and CFHT broad band systems. We have flagged all measurements where the overlap between the systems response and the unmasked spectra is below 70\%.

\begin{figure}
  \centering
  \includegraphics[width=1.\linewidth]{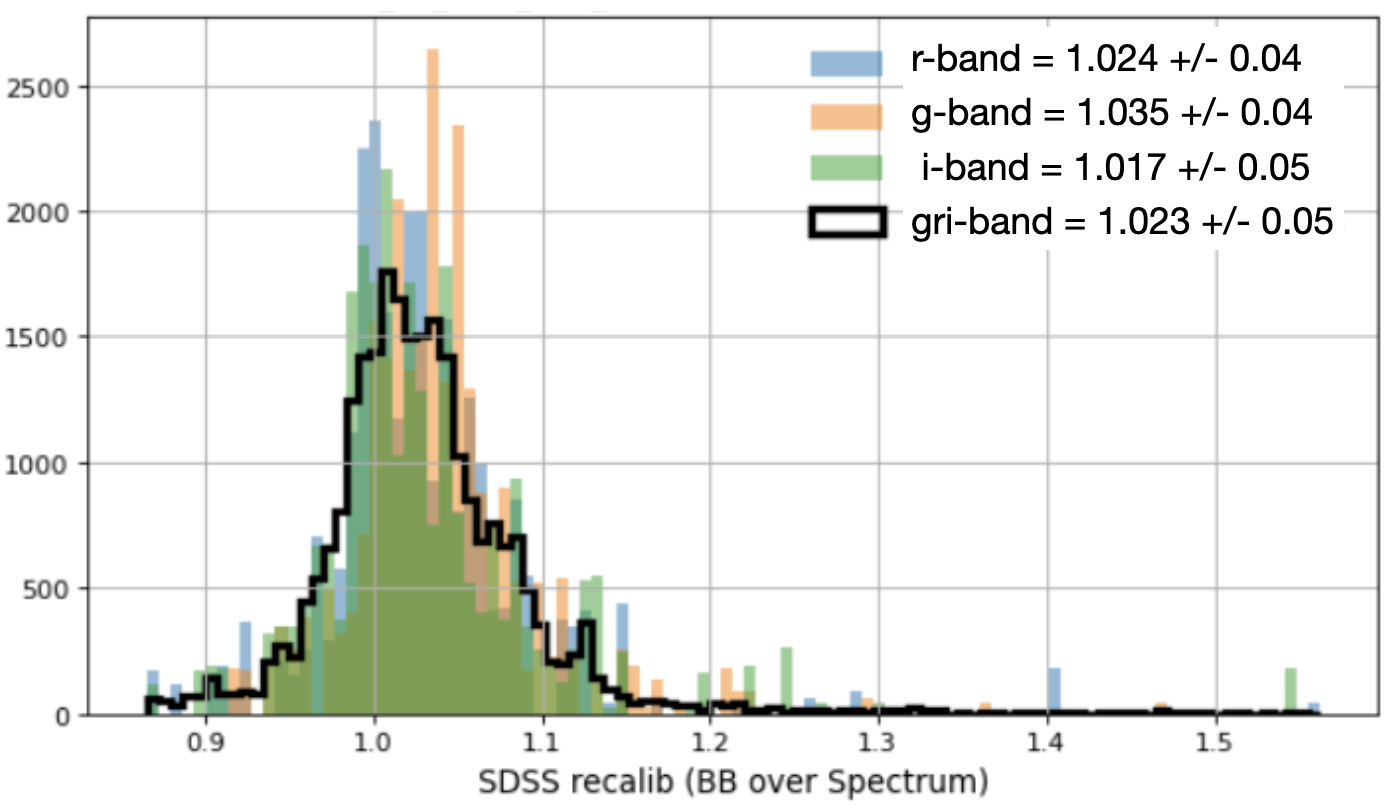}
  \caption{The coloured histograms show the values of the SDSS zero point of the spectra re-calibration ZP in Eq.\ref{eq:Rsdss} for different BB filters. The ZP values in the g, r and i bands are fit with a quadratic function of $\lambda$ to find a functional form for ZP so that it can be applied to compute the re-scaled synthetic NB fluxes. The black histogram shows the NB ZP applied using this function. The labels show the mean and $\sigma_{68}$ of the distributions.}
  \label{fig:SDSSrecal}
\end{figure}

\begin{figure*}
  \centering
  \includegraphics[width=0.49\textwidth]{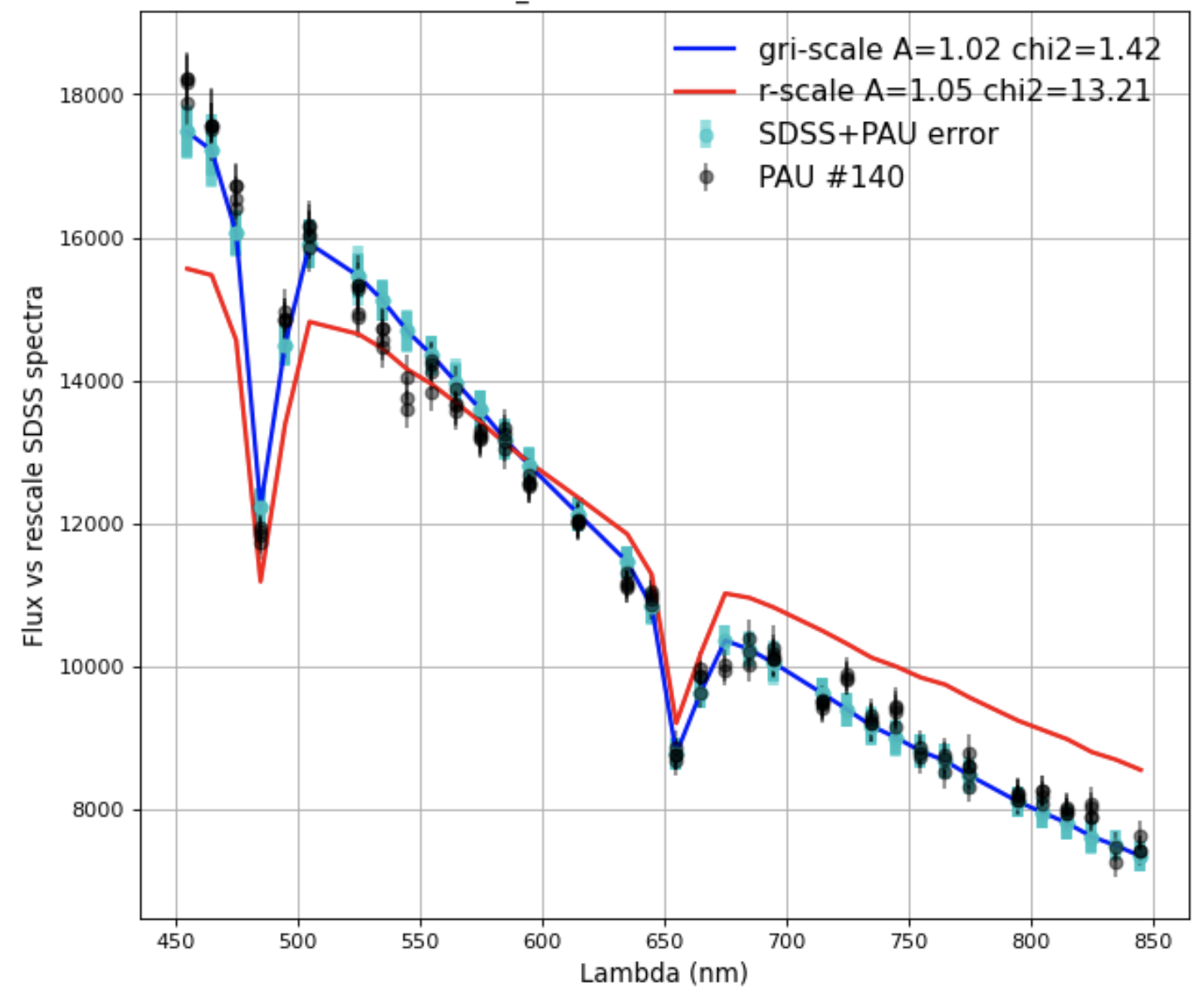}
  \includegraphics[width=0.49\textwidth]{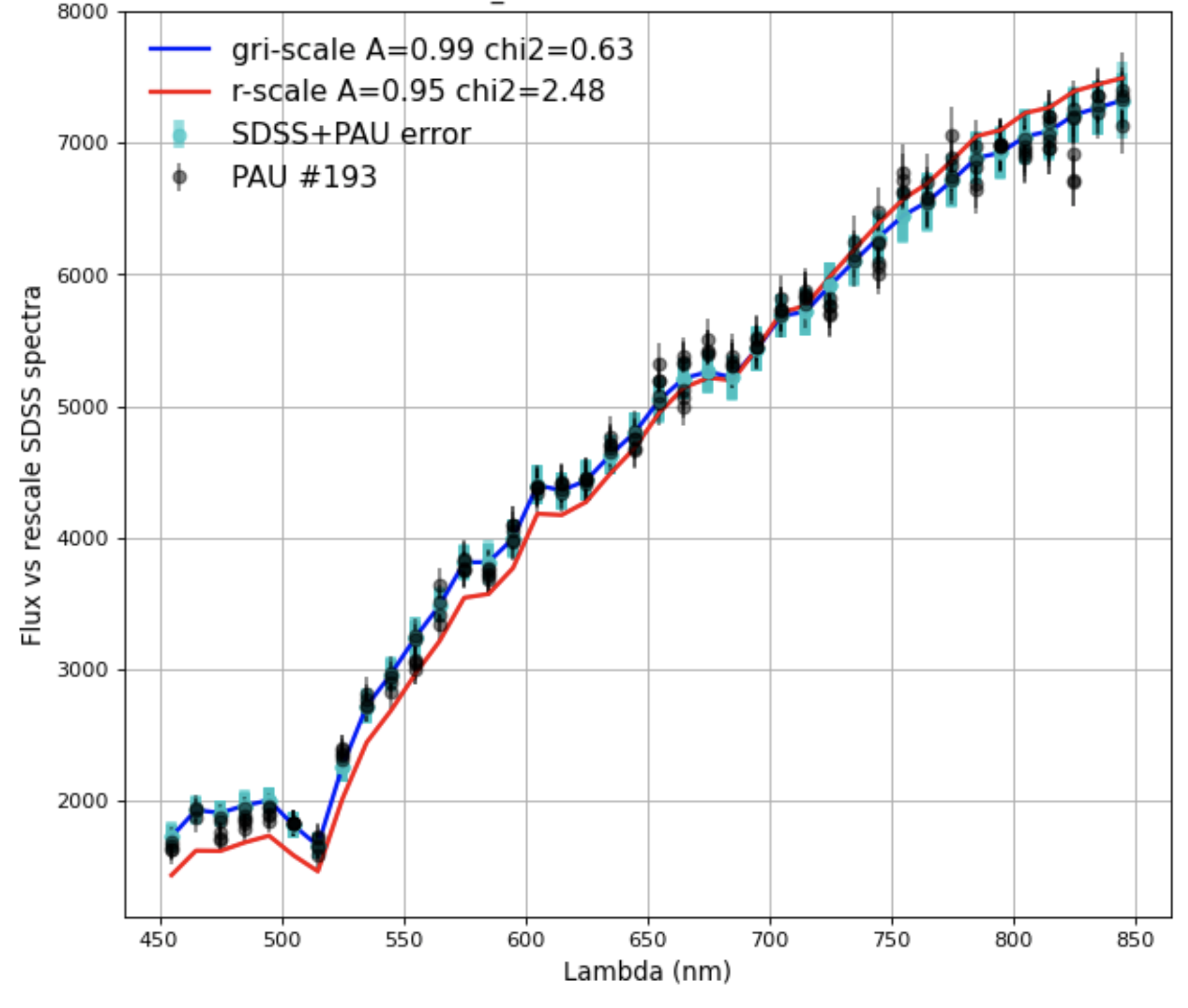}  
  \caption{Two examples of SDSS star calibration spectra for the validation study in PAUS (points with errorbars). The black points are the PAUS fluxes and the cyan points the PAUS fluxes re-calibrated with the broad band SDSS data (see section~\ref{s:recalib_SDSS_spectra}). The red lines correspond to the raw SDSS synthetic narrow band photometry re-scaled by the constant ZP value in Eq.~\ref{eq:Rsdss} to match the SDSS broad bands photometry of the same star in the $r$ band. 
  The blue lines are based on a parabolic fit to the three broad bands ($g$, $r$ and $i$) which fully overlap in wavelength with PAUS.
  In some cases (as shown in the left panel) this corrects for a colour term in the spectrum continuum.
 The spectra are also shifted by a global aperture calibration, $A$, given in Eq.~\ref{eq:Afit} (values shown in the labels), to account for possible differences in the aperture used in each observation. The "chi2" label shows the reduced $\chi^2$ as compared to the PAUS data.}
  \label{fig:validation_pau_calib} 
\end{figure*}

\subsubsection{Re-calibration of SDSS spectra}
\label{s:recalib_SDSS_spectra}

To account for remaining aperture or PSF effects in the measured SDSS spectra we use the total SDSS broad band photometry 
to re-calibrate each individual spectrum.
To do this, we first estimate the synthetic broad band fluxes from the spectra, $f_{\rm syn}^{\rm BB}$, as shown in the previous section. We then use the BB measured flux in the imaging data $f_{\rm obs}^{\rm BB}$ to find a multiplicative zero point, ZP$^{\rm BB}$, to bring both systems together. ZP$^{\rm BB}$ is in general different for each broad band filter:
\begin{equation}
{\rm ZP}^{\rm BB} = \frac{f_{\rm syn}^{\rm BB}}{f_{\rm obs}^{\rm BB}} \;.
\label{eq:Rsdss}
\end{equation}
This is equivalent to equation~\ref{eq:ZP_star_b}, where the flux coming from the combination of all the XSL stellar templates, $\langle f_{\rm star}^{\rm b} \rangle$ is now the synthetic flux, $f_{\rm syn}^{\rm BB}$.
We use the ${\rm ZP}^{\rm BB}$ values to re-scale each individual spectrum. When there are 2 (or 3) BB measurements fully within the spectral wavelength coverage, we combine them using a fit to a linear (or quadratic) function ${\rm ZP}={\rm ZP}^{\rm BB}(\lambda)$, where $\lambda$ is the mean wavelength of the bandpass response $R(\lambda)$. Each synthetic NB flux estimated from the spectrum with mean wavelength $\lambda_{NB}$ is re-scaled by the ZP function at that wavelength: ${\rm ZP}^{\rm NB}={\rm ZP}(\lambda_{\rm NB})$. 
Figure~\ref{fig:SDSSrecal} shows the histogram of values of ZP$^{\rm NB}$ for all 40 NBs in 25,644 independent measurements of 170 different SDSS calibration stars. The mean re-calibration is only 2\% with a 4\% scatter. 
Figure~\ref{fig:validation_pau_calib} shows two typical examples of SDSS synthetic spectra (with two different re-calibrations) together with the PAUS NB measurements.

\subsubsection{Aperture corrections: global amplitude}

Once the SDSS spectra are re-calibrated with the ZP coming from Eq.~\ref{eq:Rsdss}, we also perform an aperture correction of the amplitude of each individual spectrum $(S)$ to match the PAUS measurements. This is a fit to a linear constant $A=A(S)$ 

\begin{equation}
    A(S) = \frac{ \sum_i f_{\rm PAUS}(S,i) \, f_{\rm SDSS}(S,i) \, w_i} { \sum_i f_{\rm SDSS}^2(S,i) \, w_i}     
\label{eq:Afit}
\end{equation}
between the raw PAUS fluxes $f_{\rm PAUS}$ and the SDSS re-scaled synthetic spectral fluxes $f_{\rm SDSS}$ (including the spectral recalibration). The sum is over individual PAUS measurements $i$ in a given spectra $(S)$ and it
uses inverse variance weighting $w_i=1/\sigma_i^2$, where $\sigma_i$ is the joint error (from SDSS and PAUS) added in quadrature.
Typically there are 200 independent PAUS measurements (40 NB times 5 exposures) for each SDSS spectrum. 

\begin{figure}
  \centering
  \includegraphics[width=1\linewidth]{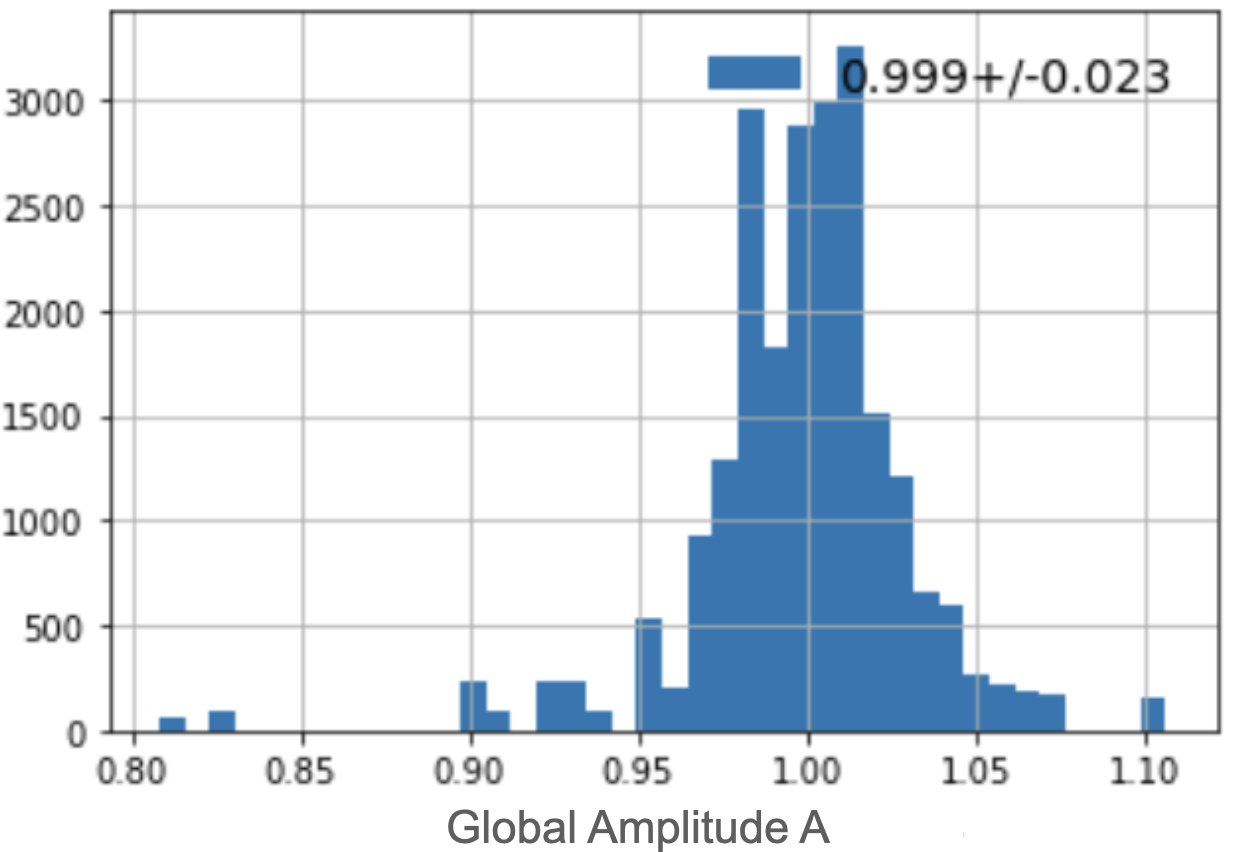}
  \caption{Aperture corrections $A$ relative to PAUS data for $N=194$ independent SDSS star calibration spectra and $42\,420$ independent PAUS measurements in COSMOS. There is a very good overall agreement. 
  We find an average factor $A= 0.999 \pm 0.023$.}
  \label{fig:SDSSrecal2}
\end{figure}

Figure~\ref{fig:SDSSrecal2} shows the distribution of values of $A$ for $N=194$ different SDSS star calibration spectra and 42\,420 independent PAUS measurements 
in COSMOS. We find a mean value and scatter of $A=0.999 \pm 0.023$, which indicates that the PAUS data are overall in very good agreement  with the SDSS calibration within 2\% overall scatter.

\begin{figure*}
  \centering
  \includegraphics[width=.8\textwidth]{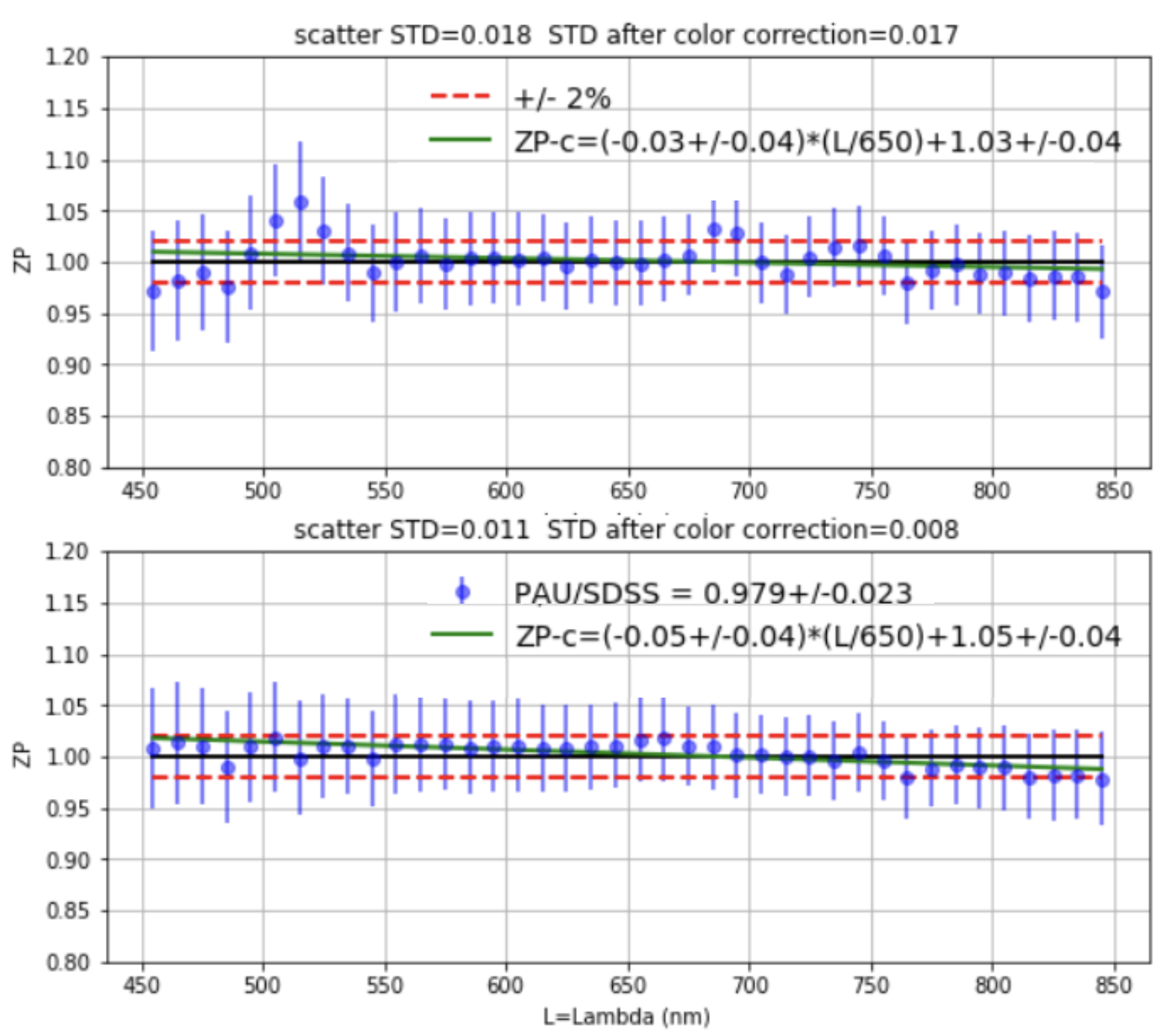}
  \caption{Validation study of the photometric calibration in PAU using synthetic narrow band photometry from SDSS spectra. The bottom panel only uses blue stars (with $g-r<0.4$) to  estimate the image zero point, while the top panel uses all stars.  
  This result demonstrates the band to band photometry is accurate relative to SDSS down to 0.8\% after correcting for a small relative tilt via Eq.\ref{eq:tilt}.}
  \label{fig:validation_pau_calib_sdss_star} 
\end{figure*}

\subsubsection{Colour terms}

We now check for any residual differences in the PAUS measured NB fluxes and the SDSS spectra synthetic values as a function of NB wavelength $\lambda$.

Figure~\ref{fig:validation_pau_calib_sdss_star} shows the mean and scatter of the zero point ratio, ZP($\lambda$), for each NB, ($\lambda$), between the PAUS raw fluxes $f_{\text{PAUS}}$ and the SDSS re-scaled synthetic spectral fluxes $f_{\text{SDSS}}$ (including the aperture correction $A$ in Eq.~\ref{eq:Afit})
\begin{equation}
    {\rm ZP}(\lambda) = \frac{ \sum_j f_{\text{PAUS}}(\lambda,j) \, f_{\text{SDSS}}(\lambda) \, w_j} { \sum_j f_{\text{SDSS}}^2(\lambda,j) \, w_j} \;,  
\end{equation}
where the $j$-index sum is over all individual PAUS measurements (42\,420 in total) and
uses inverse variance weighting $w_j \equiv 1/\sigma_j^2$, where $\sigma_j$ is the joint error added in quadrature.
There is a very good agreement overall as a function of $\lambda$. The typical scatter between the 40 NB around unity is only 1.1 to 1.8\%, depending on the star selection used to estimate the image zero point in Eq.\ref{eq:zpimage}.

The top panel in Fig.~\ref{fig:validation_pau_calib_sdss_star}
uses all stars to calibrate the image zero points. This gives the $A$ distribution shown in Fig.~\ref{fig:SDSSrecal2} with $A=0.999 \pm 0.023$.
Although the overall agreement is good, it shows some significant variations in some particular PAUS narrow bands. They correspond to line absorptions (such as the MgI triplet at $\sim5175$ {\AA}) with variations that cannot be resolved by the SDSS broad band photometry and our stellar template set and produce small biases in the corresponding narrow-bands zero-point fitting. 

The bottom panel only uses blue stars (with $g-r>0.4$) to  estimate each image zero point (see section~\ref{s:blue_stars}).
Such blue stars do not have strong absorption or emission lines and are therefore better suited for our calibration method to infer narrow band photometry from broad band photometry. This results in $A=0.979 \pm 0.023$.
In both panels, we use the same SDSS spectra and PAUS NB data (but with different image, $\text{ZP}_{\rm image}^{\rm b}$). Clearly using only blue stars (for the  image zero point of Eq.~\ref{eq:zpimage})
produces a much better agreement between the SDSS spectra and the PAUS NB. 
The overall scatter reduces from 1.8\%  to 1.1\%.

\begin{figure*}
  \centering
  \includegraphics[width=.49\textwidth]{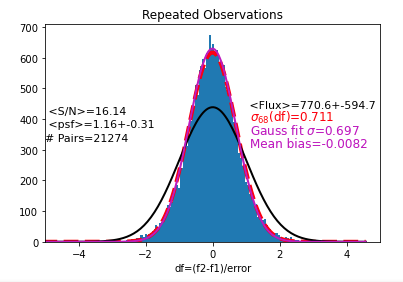}
    \includegraphics[width=.49\textwidth]{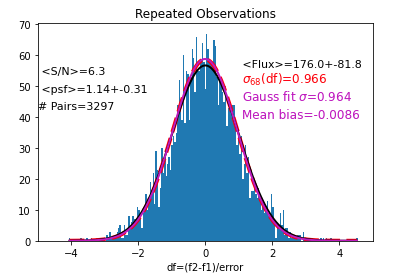}
  \includegraphics[width=.49\textwidth]{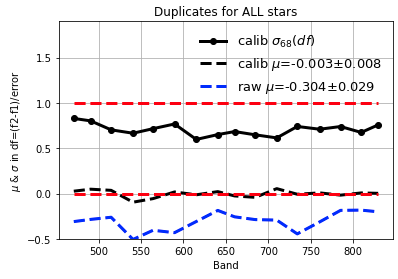}
  \includegraphics[width=.49\textwidth]{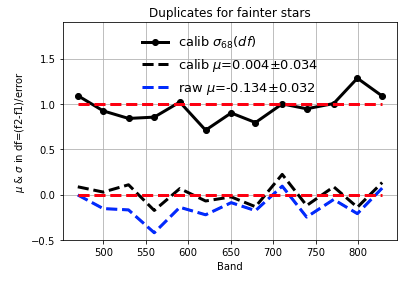}  
  \caption{Statistics of duplicate PAUS flux measurements of the stars used for calibration for all the NB. We show histograms of values of  $df \equiv (f_{\rm calib}^2 - f_{\rm calib}^1)/\sigma_{f_{\rm calib}}$, where $f_{\rm calib}^1$ and $f_{\rm calib}^2$ are the two flux measurements, with $f_{\rm calib}^2$ being the measurement in the duplicate pair with lower transparency and $\sigma_{f_{\rm calib}}$ is the joint flux error added in quadrature. 
  The top left panel corresponds to all calibration stars. The top right panel shows only pairs with faint fluxes $f_{\rm calib}^1+f_{\rm calib}^2<500$ ${\rm e}^{-}/s$. The black line corresponds to the Gaussian expectation with zero mean and unit variance. The red dashed line shows a Gaussian fit to the actual measurements.
  Bottom: mean bias, $\mu$ (dashed line), and standard deviation, $\sigma$ (solid line), of the histogram of $df$ values as a function of NB for the calibration stars in the panel above. The blue dashed line corresponds to the raw uncalibrated measurements.}
  \label{fig:duplicates} 
\end{figure*}

There seems to be a small residual colour tilt between the SDSS and PAUS NB systems.
If we fit a linear colour term to the bottom panel we find:
\begin{equation}
    {\rm ZP}(\lambda) =  1.05 \pm 0.04 - (0.05 \pm 0.04) \left(\frac{\lambda}{6500 \,\text{\AA}}\right) 
\label{eq:tilt}    
\end{equation}
which is consistent with unity within errors, so is not very significant.
The scatter between the 40 bands after correcting for this linear residual slope is just 0.8\% (1.7\% for the top panel of Fig.~\ref{fig:validation_pau_calib_sdss_star}). The scatter increases from 0.8\% to 1.1\% without the linear colour correction (Fig.~\ref{fig:validation_pau_calib_sdss_star} bottom panel).

 A similar validation using fainter spectra from SDSS and VIPERS\footnote{\url{http://vipers.inaf.it/}}~\citep{Guzzo:14,Scodeggio:18} galaxies is presented in Figure 26 of the PAUS Data Reduction paper~\citep{Serrano:2022}. That figure is done using the W1 PAUS field, instead of the COSMOS field.
 In that comparison, we also find a very good absolute and relative calibration of the PAUS NB photometry compared to VIPERS as a function of wavelength. But such validation is subject to uncertainties due to larger variations because of larger aperture effects for extended objects and also colour gradients within the different apertures used.
 We have also done similar validations for the PAUS wide fields (W1, W2 and W3, see Fig.~\ref{fig:sdss_stars} for their location in the sky) and find very similar results for both stars and galaxies.

\subsection{Duplicate observations test}
\label{subsec:duplicates}

The PAU Survey observational strategy is designed to sample the same region of the sky with several exposures with each filter. Most of these exposures of the same region of the sky with the same filter are taken in different nights.\footnote{The first two runs of the PAUS survey had observations of the same field taken consequently, but we changed the strategy after those runs. More than 90\% of the images were taken in different nights.} The tiling of the fields also includes some overlap between adjacent tiles.
We can use the repeated exposures over the PAUS fields to validate the calibration.
We have between 3-10 independent flux measurements for the same object in each NB. Most of them can be considered independent as they were taken in different nights.
In this test we only use calibration stars that are classified as stars in both the SDSS and COSMOS catalogues.
We use these catalogues to build a sample of pairs of duplicate (repeated) measurements of the same calibration star under different observing conditions.
The goal is to test if
\begin{enumerate}
    \item there is a bias in the calibrated fluxes;
    \item uncertainties in the calibrated fluxes are consistent with repeated measurements for the same object. 
\end{enumerate}
We test this as a function of the different properties of the object and observing conditions. We focus on fainter calibration stars that have fluxes closer to the galaxies we want to calibrate in the PAUS galaxy samples. We note that we use the \texttt{SExtractor} software to measure the fluxes of the calibration stars, but we use \textsc{MEMBA} (the PAUS forced aperture photometry software) for galaxies.
The \textsc{MEMBA} photometry is described in the PAUS data reduction paper~\citep{Serrano:2022} where tests of its performance are presented.

The statistics of the duplicate measurements are shown in Fig.~\ref{fig:duplicates}.
The top histograms show the values of 
\begin{equation}
 df_{\rm calib} \equiv (f_{\rm calib}^2- f_{\rm calib}^1)/\sigma_{f_{\rm calib}},  
 \label{eq:df}
\end{equation}
where $f_{\rm calib}^1$ and $f_{\rm calib}^2$ are the individual calibrated flux measurements (equation~\ref{eq:f_calib}) for the duplicate pair,
 $f_{\rm calib}^2$ is the measurement taken in lower transparency conditions~\footnote{We estimate the transparency of an observation comparing the observed flux to the SDSS flux for stars in common in the 8 central CCDs.} and $\sigma_{f_{\rm calib}}$ is the flux error from $f_{\rm calib}^1$ and $f_{\rm calib}^2$ added in quadrature. The $\sigma_{68}$ values\footnote{The $\sigma_{68}$ is defined as the half width of the probability distribution function (PDF) containing 68\% of the probability, that is, $\sigma_{68} = 0.5 ({\rm PDF}^{86} - {\rm PDF}^{16})$.} of the histograms are 
 $\sigma_{68} \simeq 0.70$ 
 for all stars and $\sigma_{68} \simeq 0.97$ for the fainter ones, that we select choosing approximately the 10\% fainter stars which  corresponds to $f_{\rm calib}^1+f_{\rm calib}^2<500$ ${\rm e}^{-}/s$. The standard deviation value increases towards $\sigma_{68}=1$ as we select fainter magnitudes.
The best fit Gaussian function to the histogram (red dashed line)  has
 mean $\mu \simeq -0.008$ and $\sigma \simeq 0.70$ for all stars and $\mu \simeq -0.009$ and $\sigma \simeq 0.97$ for the faintest ones. The bottom panels in 
Fig.~\ref{fig:duplicates} show $\mu$ and $\sigma$ as a function of NB wavelength for all stars (left) and the faintest stars (right). The results are somewhat noisier because there are fewer measurements per NB, but we can see that there is no strong tendency as a function of wavelength. 

We also plot (as blue dashed lines) the mean bias $\mu$ for the uncalibrated fluxes.
These are systematically negative because $f_{\rm calib}^1$ is always the duplicate in the pair with larger transparency and this results in negative $df_{\rm calib}$ in Eq.~\ref{eq:df}, but this bias is corrected well with the ZP calibration: for all stars the raw bias is $\mu = -0.304 \pm 0.029$ and
the calibration reduces this relative bias to $\mu = -0.003 \pm 0.008$. The correction is smaller for the fainter stars because the errors are larger. But this is still a very noticeable effect that validates the calibration process.

We also find that the calibrated flux errors (from combining the \texttt{SExtractor} errors and the ZP errors as in Eq.~\ref{eq:caliberror}) are overestimated
for all the stars (as $\sigma<1$) while they are about right for the faintest stars.
When we perform a similar duplicate pairs comparison for galaxies using the MEMBA fluxes we find that analogously the errors for bright galaxies are underestimated while errors for the fainter galaxies ($I_{\text{auto}}<21$) are  about right~\citep{Serrano:2022}. The errors for brighter fluxes are more difficult to estimate because the statistical component is very small and the errors are dominated by systematic variations across the exposure, which produce correlated error bars. This makes error propagation more difficult and less accurate. 
For the fainter fluxes, of most interest for the PAU Survey, we find very good agreement in the duplicates analysis, with small bias ($\mu \simeq 0.4\%$) and consistent errors within $3\%$ accuracy ($\sigma \simeq 0.97$).

In general we find that the mean bias $\mu$ is very small, below 0.5\%, relative to $\sigma_{f_{\rm calib}^{\rm b}}$. This is a good indication that the calibration is working well on average. 
Notice that $\mu$, even when very small, is negative, which indicates that the calibration of fluxes taken in conditions with higher transparency make the calibrated fluxes slightly biased high with respect to the ones taken with lower transparency conditions.

\begin{figure*}
    \centering
    \includegraphics[width=.49\textwidth]{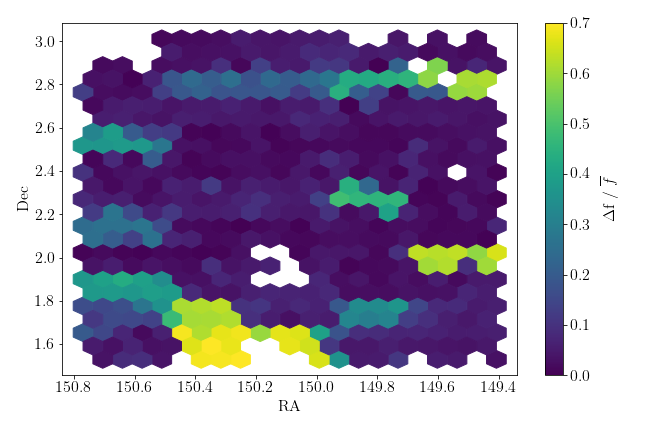}
    \includegraphics[width=.49\textwidth]{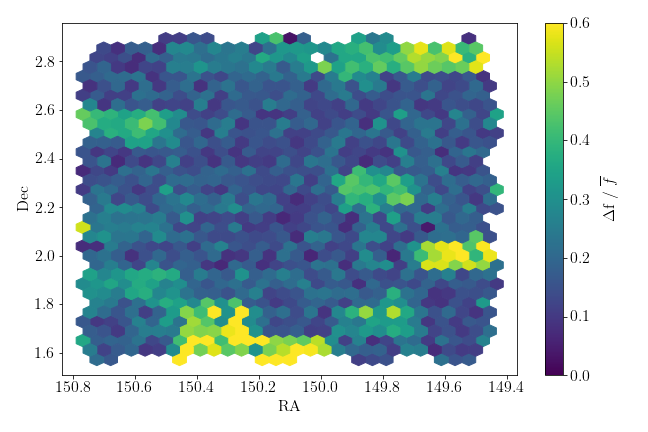}
    \includegraphics[width=.49\textwidth]{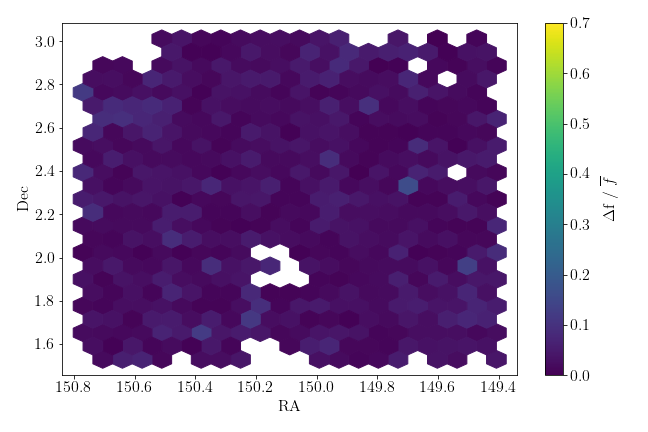}
    \includegraphics[width=.49\textwidth]{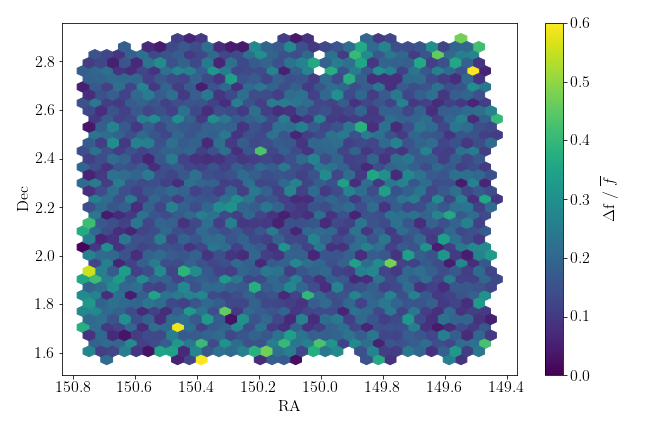}
\caption{Duplicate PAUS flux measurements over the COSMOS field in the NB585 band for calibration stars (left) and galaxies (right). The top panels illustrate the flux measured prior to the photometric calibration while the bottom panels include the image zero point calibration, drastically reducing the difference between overlapping exposures.}
\label{fig:spatial_df}
\end{figure*}

Additionally, we have evaluated the spatial dependence of the uncalibrated and calibrated flux residuals between overlapping sources. This is particularly important in any cosmic survey where spatially dependent calibration may result into undesired bias of the scientific result. Figure~\ref{fig:spatial_df} illustrates a validation test over the COSMOS field. The areas with larger difference correspond to observations with higher atmospheric extinction and the shape of these patterns resemble the area of a single detector image.
For both the calibration stars and the galaxy aperture fluxes (left and right panels in the figure, respectively) the duplicate difference is dramatically reduced after applying the image zero point to the corresponding measurement (top to bottom panels difference).

\subsection{Spectrophotometric standards}
\label{subsec:spectrophotometric}

Another method to check the calibration of the PAU Survey is to use a sample of spectrophotometric standard stars that covers the spectral range of the PAUS narrow band filters, from 4500 to 8500 {\AA} and with a distribution on the sky that allows us to observe them in the different runs assigned during the year. We have selected a subsample of the \citet{standards-oke1990} spectrophotometric standard stars. Their spectra can be accessed from the ESO webpage\footnote{\url{https://www.eso.org/sci/observing/tools/standards/spectra/stanlis.html}}.

\subsubsection{Observations}
Seven spectrophotometric standard stars were regularly observed during the PAU survey observation runs. These stars are listed in Table~\ref{tab:StandardStars}. Since they are relatively bright, with V-band magnitudes around 10--12, we took shorter exposures than for regular PAUS observations\footnote{Regular PAUS observations range between 80 and 200 seconds depending on filter tray and sky conditions.}, ranging between 10 and 15 seconds depending of the star brightness and the filter tray.
These standard stars were imaged in both the evening and morning twilights.

Since 2015, we have obtained 6,460 images of these spectrophotometric standards stars. Table \ref{tab:StandardStars} lists the amount of exposures of each standard star and each filter tray.

\begin{table*}
    \begin{center}
        \begin{tabular}{| l | c | c | c | c | c | r|}
          & NB455-525 & NB535-605 & NB 615-685 & NB695-765 & NB775-845 & Total\\
         \hline     
         BD254655  &    49     &    80     &     96     &    96     &    64     &   385\\
         BD284211  &    48     &    72     &     80     &    88     &   119     &   407\\
         BD332642  &   198     &   198     &    301     &   220     &   175     & 1,092\\
         BD75325   &    72     &   128     &    152     &   168     &   176     &   696\\
         Feige 34  &   413     &   429     &    459     &   489     &   566     & 2,356\\
         Feige 110 &   131     &   184     &    208     &   261     &   316     & 1,100\\
         G191 B2B  &    80     &    56     &    112     &    88     &    99     &   424\\
         \hline
         Total     &   991     & 1,147     &  1,408     & 1,410     & 1,504     & 6,460\\
         \hline
        \end{tabular}
        \caption{List of spectrophotometric standard stars used in PAUS since 2015 with the number of observations for each filter tray.}
        \label{tab:StandardStars}
    \end{center}
\end{table*}

\subsubsection{Measurements}
The fluxes for each standard star and narrow band filter are obtained with the same pipeline used to analyse the survey data described in~\citet{Serrano:2022}. Due to the short exposure time of these images, the number of stars of the SDSS catalogue that are bright enough to be detected at high signal-to-noise ratios to be used for calibration is extremely low. For this reason we cannot compute a zero point based on them as in the longer exposure survey images (as in section~\ref{section: Photometric calibration}).

The fluxes are converted to magnitudes and corrected for atmospheric extinction using the values provided by King (1985)
for the Roque de los Muchachos Observatory (see footnote~\ref{footnote:technote}).

The observed magnitudes for each image are compared to the synthetic magnitudes obtained from the spectrophotometric standard spectra taking into account the NB filter transmission curve as in sections~\ref{s:synth_pau_mags} and~\ref{s:synthetic_photometry}. The zero point  for each observation is evaluated by subtracting the computed synthetic magnitude from the observed narrow band filter magnitude corrected for atmospheric extinction.

The global ZP for each narrow band filter is taken as the median value of the ZP obtained from each observation of that NB filter. Its error is taken as the standard deviation associated with the Median Absolute Deviation (MAD) $(\sigma = 1.4826\cdot MAD)$.

\subsubsection{Comparison with the main calibration}
We want to compare the ZP obtained with the spectroscopic standards to the ZP calculated for the PAUS images. As described in~\cite{Serrano:2022} and in the previous sections, the zero points in the PAUS survey are evaluated comparing observed fluxes to synthesised fluxes for SDSS stars. These ZPs are used as multiplicative factors to the observed fluxes in the PAUS data management system. Traditionally, the ZP are given as additive terms to the magnitudes as they can be easily interpreted as the magnitude of an object that produces an observed flux of 1 ${\rm e}^{-}/s$. Given that the use of spectrophotometric standards is closer to traditional calibration methods we have decided to keep the ZP computed this way as an additive term. Therefore, in order to compare the previous PAUS multiplicative zero points (${\rm ZP}_{\times}$), we convert them to additive terms (${\rm ZP}_{+}$)  
\begin{equation}
    {\rm ZP}_{+} = -2.5\cdot \log _{10} ({\rm ZP}_{\times})   
\label{eq:ZP+ZPx}
\end{equation}

In the PAU Survey, the computed zero points contain the atmospheric extinction term (ZP$_1$ in Eq.~\ref{eq_ZP}) while the zero points calculated with the spectroscopic standards do not contain it (ZP$_2$ in Eq.~\ref{eq_ZP}). Therefore, we apply an atmospheric correction to the additive PAUS survey ZP to convert them from the ZP$_1$ to the ZP$_2$ type. Equally to what we did with the spectroscopic standards, we use the extinction coefficients of King (1985) and the value of the airmass of the observations for the atmospheric correction. Once corrected to zero points above the atmosphere, we median combine the zero points of all the observations with good transparency according to the PAU data management reduction \citep{Serrano:2022} in each filter to obtain a global filter ZP to compare with the one obtained from the spectrophotometric standards. 

\begin{figure}
  \centering
  \includegraphics[width=\linewidth]{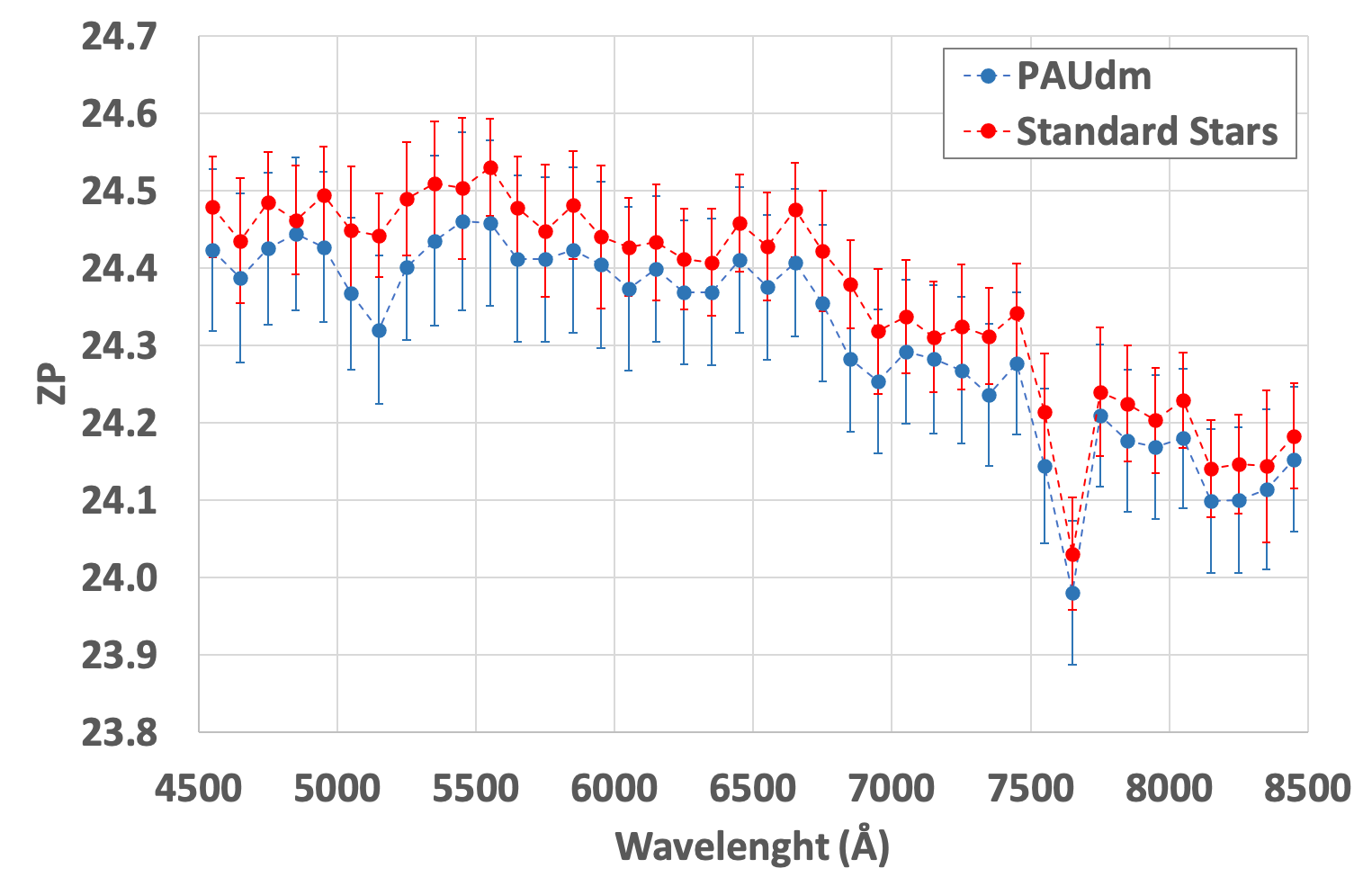}
  \caption{Additive zero point, ${\rm ZP}_{+}$, computed with the spectrophotometric stars (red) and coming from the calibration of the PAUS survey images with SDSS stars (blue).}
  \label{fig:zero_points_comp}
\end{figure}

\begin{figure}
  \centering
  \includegraphics[width=\linewidth]{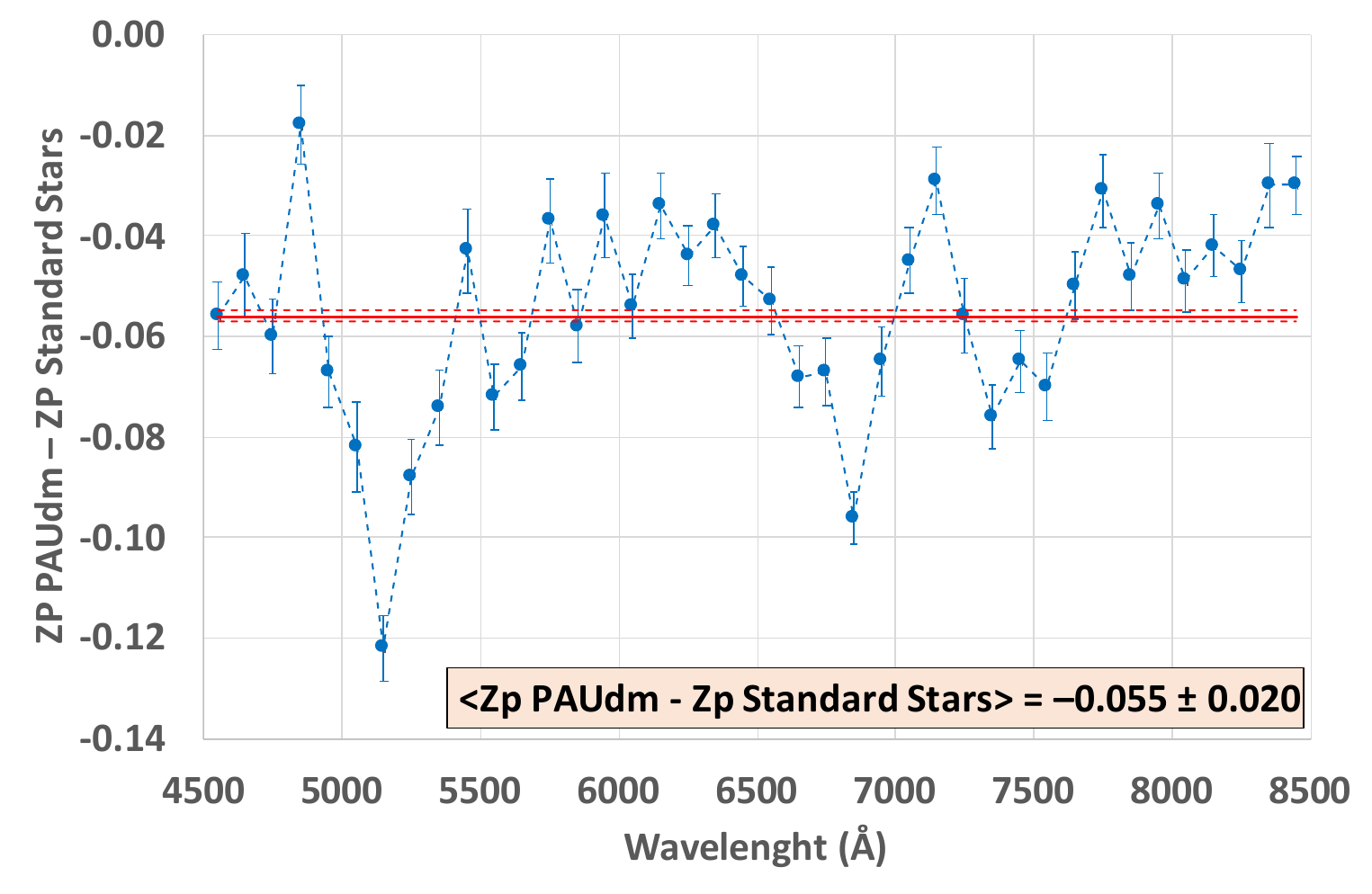}
  \caption{Difference of the additive zero points determined using the PAUS data management system and the spectrophotometric stars as a function of NB filter (or wavelength).}
  \label{fig:zero_points_diff}
\end{figure}

Figure~\ref{fig:zero_points_comp} shows the comparison of the values of the zero points computed using spectrophotometric standard stars in red and the values obtained in the PAUS survey with our calibration method in blue. There is a small overall shift between both calibrations. Figure~\ref{fig:zero_points_diff} shows the difference of the zero points as a function of NB filter (or wavelength). The mean value of the difference is $\Delta {\rm ZP} = -0.055 \pm 0.020$ and there is no apparent wavelength trend. The overall difference may be due to the fact that we have used zero points from images taken with a variety of transparencies and there may be a grey term to the atmospheric extinction not taken into account. Otherwise, it can mean a different calibration of the spectrophotometric standards compared to the SDSS stars used for the PAUS calibration. In fact, the SDSS star calibration is known to be consistent with the AB system only at the one percent level in the $g$, $r$ and $i$ filters\footnote{\url{https://www.sdss.org/dr12/algorithms/fluxcal/}} (and somewhat larger in the $u$ and $z$ filters), which is not enough to explain this difference.
In Fig.~\ref{fig:zero_points_diff} we can appreciate two features we already found in Fig.~\ref{fig:star_image_zp_blue} when comparing the zero points obtained with the bluest stars only to the ones obtained with all the stars. The MgI triplet at $\sim5175$ {\AA} and the telluric B absorption at $\sim6850$ {\AA} features yield a brighter value of the zero point which increases the median of the calibration offset.

Overall, this comparison test to spectrophotometric standards reinforces the view that the PAUS calibration does not have any significant wavelength dependence deviation compared to this spectrophotometric standard system.

\section{Summary and Conclusions}
\label{section: Summary and Conclusions}

The PAU Survey is imaging large areas of the sky (currently $\sim 50 \,{\rm deg}^2$) with the PAU Camera using narrow band filters. The filter set is composed of forty 130 {\AA} wide filters designed to be contiguous in wavelength coverage with central wavelengths separated by 100 {\AA} and spanning the wavelength range from 4500 to 8500 {\AA}. The PAU survey has currently published a series of papers exploiting the data taken.
The scientific analysis described in these papers relies in one way or another on the photometric calibration of the narrow band data. 

The calibration of the PAUS data relies on the photometry of the SDSS. In each PAUS exposure we select stars that are measured in the SDSS. We compare these stars to the XSL spectral library.
We compute synthetic fluxes for all the XSL templates in the SDSS and PAUS filter systems for a variety of the Milky Way extinction values. We generate a synthetic spectrum for each SDSS star in common with PAUS combining  the XSL templates weighted by their goodness of fit to the SDSS broad band measurements. We use these synthetic templates to compute the expected fluxes 
in the PAUS NB filters. We compute a zero point for each star in common for all the PAUS filters comparing the observed fluxes to the synthetic ones. We obtain the zero point of each exposure in each filter by combining the zero point of all the individual stars. 

We check the reliability of the calibration. We compare the individual star zero points to the combined image zero points. We test the effects due to the background subtraction finding no significant trends as a function of incoming flux. We check the influence of the choice of stars in the calibration. 
We compare the calibration zero points obtained from only blue stars to the calibration obtained from all the stars. We find the calibrations to be consistent overall at the 2\% level except some differences around $\sim 5170$ {\AA} due to the MgI triplet stellar absorption and also redder than $\sim 6850$ {\AA} where the atmospheric telluric absorption and emission lines dominate the sky spectrum (Fig.~\ref{fig:star_image_zp_blue}). 
 
The PAU Camera field of view suffers from vignetting at the edges of the eight central CCDs where the NB filters are located. In the outermost central CCDs the illumination variation can reach 8\% peak-to-valley (Fig.~\ref{fig:star_image_zp_spatial}). 
The statistical comparison of the individual star ZPs to the total image ZP as a function of detector position serves to build a star flat field that can correct the illumination pattern variations. 
 
We also checked our calibration using SDSS stellar spectra instead of the SDSS stellar photometry. 
First we recalibrate each individual SDSS spectrum using its SDSS photometry. We then compute an overall factor between the all the PAUS measurements and the synthetic fluxes obtained from the SDSS spectra already rescaled to the SDSS photometry. 
The agreement is very good assessing our process of fitting stellar templates from the SDSS broad band photometry to compute the synthetic NB photometry is not changing the overall PAUS calibration.
When studying the calibration as a function of wavelength, we have considered two cases: taking into account all the stars (Fig.~\ref{fig:validation_pau_calib} top) and only considering the blue stars (Fig.~\ref{fig:validation_pau_calib} bottom). In the later case, the calibration does not show any significant filter/wavelength trend above the 1\% level. This is not the case when we use all the stars in the PAUS calibration, obtaining the same calibration error features as with the individual stars to total image ZP ratio study (Fig.~\ref{fig:star_image_zp_blue}). 

Next, we check how robust our calibration is comparing repeated observations of the same stars that have undergone a different calibration as the calibration is done per exposure. We find that the mean relative difference once the duplicates are calibrated is consistent with zero (dashed black line in bottom panels of Fig.~\ref{fig:duplicates}) which is not the case when the fluxes are uncalibrated (dashed blue line in bottom panels of Fig.~\ref{fig:duplicates}). So, our calibration is consistent for different measurements of the same objects. We also find our errors to be somewhat underestimated (dashed black line in left bottom panel of Fig.~\ref{fig:duplicates}). The errors are better estimated when we consider faint sources (dashed black line in right bottom panel of Fig.~\ref{fig:duplicates}). In addition, in \citet{Serrano:2022} we also analysed the duplicate measurements of galaxies, finding consistent calibration throughout different observations taken with different atmospheric conditions.

Finally, we check our calibration with the one obtained with spectrophotometric standards. We find that both calibrations are consistent as a function of filter/wavelength except a small offset at the 5\% level and a couple of small features (Fig.~\ref{fig:zero_points_diff}). The comparison is not straight forward as it depends on the atmospheric extinction that we have not calibrated. Instead we have assumed the extinction coefficients of King (1985). Moreover, although we have used only PAUS images with relatively high transparency, there are nevertheless images with lower values of the transparency which can explain the small offset ($\Delta ZP \sim 0.05$ in magnitudes) we find between both calibrations. We also find the same $\sim$5175 {\AA} MgI  and $\sim$6850 {\AA} B band telluric absorption features where the calibrations show a higher discrepancy. 

Lastly, although not discussed in this paper, the photometric redshifts that we infer from our data \citep{Eriksen:2019,Eriksen:2020,Alarcon:2021,Soo:2021} are consistent with the ones we predicted in our simulations where perfect calibrations where assumed~\citep{Marti:2014}. This is an indication that our calibration probably does not have any  significant calibration inaccuracy as a function of filter or wavelength. In addition, template fitting photometric codes can be run to check the relative offsets between bands when the spectroscopic redshifts are known. We do not find any significant offset when we apply this test with the PAUS data. 

Overall, based on the tests performed, we validate our calibration procedure. We assess that our photometric calibration is close to the AB system with relative errors at the 2\% level when we calibrate with all the stars and around 1\% when only using blue stars. Any trend as a function of wavelength is below a 1\% change. Our calibration procedure may well serve other surveys wanting to calibrate narrow or medium band data.

\section {Data availability}
The data underlying this article will be shared on reasonable request to the corresponding author.

\section*{Acknowledgements}

The PAU Survey is partially supported by MICINN under grants CSD2007-00060, AYA2015-71825, ESP2017-89838, PGC2018-094773, PGC2018-102021, PID2019-111317GB, SEV-2016-0588, SEV-2016-0597, MDM-2015-0509, CEX2020-001058-M and Juan de la Cierva fellowship and LACEGAL and EWC Marie Sklodowska-Curie grant No 734374 and no.776247 with ERDF funds from the EU Horizon 2020 Programme, some of which include ERDF funds from the European Union. IEEC and IFAE are partially funded by the CERCA and Beatriu de Pinos program of the Generalitat de Catalunya. Funding for PAUS has also been provided by Durham University (via the ERC StG DEGAS-259586), ETH Zurich, Leiden University (via ERC StG ADULT-279396 and Netherlands Organisation for Scientific Research (NWO) Vici grant 639.043.512), University College London and from the European Union's Horizon 2020 research and innovation programme under the grant agreement No 776247 EWC. HHi is supported by a German Research Foundation (DFG) Heisenberg grant (Hi 1495/5-1), the DFG Collaborative Research Center SFB1491, as well as a European Research Council (ERC) Consolidator Grant (No. 770935). PR acknowledges the support by the Tsinghua Shui Mu Scholarship, the funding of the National Key R\&D Program of China (grant no. 2018YFA040450), the National Science Foundation of China (grant no. 12073014 and 12350410365), the science research grants from the China Manned Space Project with No. CMS-CSST2021-A05, and the Tsinghua University Initiative Scientific Research Program (No. 20223080023). The PAU data center is hosted by the Port d'Informaci\'o Cient\'ifica (PIC), maintained through a collaboration of CIEMAT and IFAE, with additional support from Universitat Aut\`onoma de Barcelona and ERDF. We acknowledge the PIC services department team for their support and fruitful discussions






\bibliographystyle{mnras}
\bibliography{references} 



%
%


\bsp	
\label{lastpage}
\end{document}